\documentclass[twocolumn, tighten]{aastex6}
\usepackage[caption=false]{subfig}
\usepackage{epstopdf}
\usepackage{amsmath}
\usepackage{bookmark}
\hypersetup{citecolor=blue,linkcolor = blue,urlcolor = blue}
\begin{document}

\title{CO and dust properties in the TW Hya disk from high-resolution ALMA observations}

\author{Jane Huang\altaffilmark{1}, Sean M. Andrews\altaffilmark{1}, L. Ilsedore Cleeves\altaffilmark{1,2}, Karin I. \"Oberg\altaffilmark{1}, David J. Wilner\altaffilmark{1}, Xuening Bai\altaffilmark{1,3,4}, Til Birnstiel\altaffilmark{5}, John Carpenter\altaffilmark{6}, A. Meredith Hughes\altaffilmark{7}, Andrea Isella\altaffilmark{8}, Laura M. P\'erez\altaffilmark{9}, Luca Ricci\altaffilmark{8}, Zhaohuan Zhu\altaffilmark{10}}

\altaffiltext{1}{Harvard-Smithsonian Center for Astrophysics, 60 Garden Street, Cambridge, MA 02138, USA}
\altaffiltext{2}{Hubble Fellow}
\altaffiltext{3}{Institute for Advanced Study, Tsinghua University, Beijing 100084, People's Republic of China}
\altaffiltext{4}{Tsinghua Center for Astrophysics, Tsinghua University, Beijing 100084, People's Republic of China}
\altaffiltext{5}{University Observatory, Faculty of Physics, Ludwig-Maximilians-Universit\"at M\"unchen, Scheinerstr. 1, D-81679 Munich, Germany}
\altaffiltext{6}{Joint ALMA Observatory, Alonso de Cordova 3107 Vitacura - Santiago de Chile, Chile}
\altaffiltext{7}{Van Vleck Observatory, Astronomy Department, Wesleyan University, 96 Foss Hill Drive, Middletown, CT 06459, USA}
\altaffiltext{8}{Department of Physics and Astronomy, Rice University, 6100 Main Street, Houston, TX 77005, USA}
\altaffiltext{9}{Departamento de Astronomia, Universidad de Chile, Casilla 36-D, Santiago, Chile}
\altaffiltext{10}{Department of Physics and Astronomy, University of Nevada, Las Vegas, 4505 S. Maryland Pkwy., Las Vegas, NV 89154, USA}

\email{jane.huang@cfa.harvard.edu}

\begin{abstract}
We analyze high angular resolution ALMA observations of the TW Hya disk to place constraints on the CO and dust properties. We present new, sensitive observations of the $^{12}$CO $J = 3-2$ line at a spatial resolution of 8 AU (0\farcs14). The CO emission exhibits a bright inner core, a shoulder at $r\approx70$ AU, and a prominent break in slope at $r\approx90$ AU. Radiative transfer modeling is used to demonstrate that the emission morphology can be reasonably reproduced with a $^{12}$CO column density profile featuring a steep decrease at $r\approx15$ AU and a secondary bump peaking at $r\approx70$ AU. Similar features have been identified in observations of rarer CO isotopologues, which trace heights closer to the midplane. Substructure in the underlying gas distribution or radially varying CO depletion that affects much of the disk's vertical extent may explain the shared emission features of the main CO isotopologues. We also combine archival 1.3 mm and 870 $\mu$m continuum observations to produce a spectral index map at a spatial resolution of 2 AU. The spectral index rises sharply at the continuum emission gaps at radii of 25, 41, and 47 AU. This behavior suggests that the grains within the gaps are no larger than a few millimeters. Outside the continuum gaps, the low spectral index values of $\alpha\approx 2$ indicate either that grains up to centimeter size are present, or that the bright continuum rings are marginally optically thick at millimeter wavelengths. \end{abstract}


\section{Introduction}

Tracing the distribution of dust and volatiles in protoplanetary disks is key for guiding models of planet formation. Disk observations may be useful for constraining the formation locations of planets \citep[e.g.][]{2011ApJ...743L..16O, 2014ApJ...785..122Z, 2015ApJ...809...93D}, the mechanisms by which the dust and gas distributions in disks evolve to form planets and planetesimals \citep[e.g.][]{2012ApJ...760L..17P, 2015ApJ...813L..14B, 2015AA...579A.106V}, and the abundances of volatiles that will eventually be incorporated into planetary atmospheres \citep[e.g.][]{2016MNRAS.461.3274C, 2016AA...595A..83E,2016ApJ...831L..19O}. 

Millimeter/submillimeter interferometry plays a fundamental role in characterizing disk structures due to the high spatial resolution and sensitivity that can be achieved for observations of both dust and molecular emission. Observations of the millimeter continuum, which is dominated by thermal emission from millimeter-sized grains, can test models of grain growth and transport \citep[e.g.][]{2010AA...516L..14B,2010AA...512A..15R, 2011AA...529A.105G, 2012ApJ...744..162A}. CO is often targeted simultaneously with continuum observations; as an abundant and easily observable molecule, CO is used to infer fundamental properties such as gas masses and temperatures \citep[e.g.][]{2003AA...399..773D, 2013ApJ...774...16R, 2014ApJ...788...59W}. In addition, since CO is the primary gas-phase carbon reservoir in disks, characterizing its distribution is also pertinent to predicting the abundances and distribution of many other species that are major components of the gas and ice incorporated into planets and planetesimals \citep[e.g.][]{1999AA...351..233A, 2015AA...579A..82R}. 

Due to its proximity, relative isolation, bright emission, and nearly face-on orientation, the TW Hya disk (J2000 R.A. 11h01m51.905s, Decl. -34d42m17.03s) has long served as a template for protoplanetary disks, spurring the development of techniques, models, and lines of inquiry that have since been extended to other sources \citep[e.g.][]{2002ApJ...568.1008C, 2003AA...400L...1V, 2004ApJ...616L..11Q, 2013Sci...341..630Q, 2013Natur.493..644B}. TW Hya is a 10 Myr old K6 star in the TW Hydrae association lying 59.5 pc away from Earth \citep[e.g.][]{1997Sci...277...67K, 1999ApJ...512L..63W, 2006AA...460..695T, 2013ApJ...762..118W, 2016AA...595A...2G}.  Intriguingly, recent observations have revealed concentric rings and gaps in millimeter/submillimeter continuum emission tracing the distribution of large dust grains near the midplane \citep{2016ApJ...820L..40A, 2016ApJ...829L..35T}, scattered light observations tracing the distribution of small dust grains in the upper layers of the disk \citep[e.g.][]{2013ApJ...771...45D, 2015ApJ...802L..17A, 2015ApJ...815L..26R, 2017ApJ...837..132V}, and molecular line emission \citep[e.g.][]{2016ApJ...819L...7N, 2016ApJ...823...91S, 2017ApJ...835..228T}. The origins of these features and their relationships to one another are not yet definitively established. Embedded planets, molecular snowlines, structured magnetohydrodynamic turbulence, and photoevaporation are often invoked as hypotheses to explain the types of features observed in the TW Hya disk (see aforementioned references and, e.g., \citealt{2015AA...574A..68F, 2015ApJ...806L...7Z, 2017ApJ...843..127D, 2017MNRAS.464L..95E}). 

To place additional constraints on the dust and CO distributions in the TW Hya disk, we analyze new high angular resolution ALMA observations of the $^{12}$CO $J=3-2$ transition, as well as a spectral index map produced from archival 1.3 mm and 870 $\mu$m continuum observations. The data reduction is described in Section 2. The observed line and continuum emission are described in Section \ref{sec:results}, and radiative transfer modeling of the $^{12}$CO data is presented in Section \ref{sec:models}. We provide a discussion in Section \ref{sec:discussion} and a summary in Section \ref{sec:summary}. 

\section{Observations and Data Reduction}\label{sec:observations}

\subsection{Continuum Reduction}
We reprocessed and combined archival 1.3 mm (Band 6) and 870 $\mu$m (Band 7) continuum data from six ALMA programs. The raw data from programs 2013.1.00114.S, 2013.1.00198.S, and 2015.1.00686.S were calibrated by National Radio Astronomical Observatory (NRAO) staff, and the raw data from programs 2013.1.00387.S, 2015.A.00005.S, and 2013.1.01397.S were downloaded from the ALMA archive and calibrated in \texttt{CASA}  \citep{2007ASPC..376..127M} using the accompanying reduction scripts. Table \ref{tab:settings} summarizes the observation setups. Bright quasars were used for bandpass and phase calibration, and either a solar system object (using the Butler-JPL-Horizons 2012 model) or a bright quasar was used for amplitude calibration. Table \ref{tab:fluxcal} lists calibrator details. 

Additional calibration and imaging were performed with \texttt{CASA 4.5.3}. After flagging channels displaying strong line emission and data with anomalous amplitudes or phases, the line-free channels were spectrally averaged to form continuum visibility datasets. Data from the four programs observed with compact antenna configurations (2013.1.00114.S, 2013.1.00198.S, 2013.1.00387.S, and 2013.1.01397.S) were first individually imaged and phase self-calibrated. The continuum fluxes from the self-calibrated images were within 5\% of one another within each band (measured within a 2\farcs5 diameter region in the images to be $\approx0.58$ and $\approx1.4$ Jy for Bands 6 and 7, respectively), which is compatible with the estimated 10\% systematic flux calibration uncertainty of ALMA in these bands. Since the high-resolution datasets (2015.1.00686.S and 2015.A.00005.S) only had a few baselines at $uv$ distances below 200 k$\lambda$, the amplitudes of these visibilities were compared directly with those from the short-baseline datasets to check for consistency at corresponding $uv$ distances. Because of the relatively large proper motion of TW Hya, the \texttt{fixvis} and \texttt{fixplanets} tasks in \texttt{CASA} were used to shift the datasets to a common phasecenter.

A high-resolution 1.3 mm continuum image was produced by combining the data from the three Band 6 programs and applying the multi-scale multi-frequency synthesis algorithm \citep{2008ISTSP...2..793C}, as implemented in the \texttt{clean} task.  Briggs weighting (robust $=0.5$) and scales of  $0''$, 0$\farcs$06, 0$\farcs$15, 0$\farcs$3, and 0$\farcs$6 were used. This generated a source model used to phase self-calibrate the data together. A similar self-calibration and imaging procedure was applied to the Band 7 data. The self-calibrated 870 $\mu$m and 1.3 mm continuum images were used to check for consistency with the images published in \citet{2016ApJ...820L..40A} and \citet{2016ApJ...829L..35T}, which used the same long-baseline data and some of the same short-baseline data. The two continuum images and the deprojected, azimuthally averaged radial brightness temperature profiles are shown in Appendix B.

The self-calibrated 870 $\mu$m and 1.3 mm continuum datasets were then imaged together with the \texttt{clean} task's implementation of multi-term multi-frequency synthesis \citep{2011AA...532A..71R} with $nterms = 2$ and a Briggs robust parameter of 0. In brief, the imaging algorithm uses a first-order Taylor expansion to model the source intensity as a function of frequency, i.e.,
\begin{equation}
I_\nu = I_{\nu_0}\left(\frac{\nu}{\nu_0}\right)^\alpha\approx I_{\nu_0}\left(1+\alpha \left( \frac{\nu-\nu_0}{\nu_0}\right)\right).
\end{equation}
This procedure takes advantage of the additional $uv$ coverage offered by wide-band imaging to produce a higher-fidelity continuum image compared to imaging the Band 6 or Band 7 data individually. In addition, simultaneously fitting for the spectral index $\alpha$ during the deconvolution process also reduces the influence of imaging artifacts that would arise from computing the spectral index from the Bands 6 and 7 images individually. The combined continuum image is at a frequency of 290 GHz. The synthesized beam is $37\times26$ mas (2.2 $\times$ 1.5 AU), with a position angle of 73$\fdg$7. The rms measured in a signal-free portion of the image is 20 $\mu$Jy beam$^{-1}$. The spectral index is computed only for pixels where the intensity image exceeds 0.1 mJy beam$^{-1}$ (5 $\times$ the rms noise). 

\begin{deluxetable*}{ccccccccc}
\tabletypesize{\scriptsize}
\tablecaption{ALMA Observation Summary\label{tab:settings}}
\tablehead{
\colhead{Program} &\colhead{P.I.}& \colhead{Ref.}&\colhead{Date} &\colhead{Freq. range}&\colhead{Antennas}&\colhead{Baselines} &\colhead{On-source}&\colhead{Notes}\\
\colhead{}& \colhead{} &\colhead{}&\colhead{}& \colhead{(GHz)}&\colhead{} & \colhead{(m)} &\colhead{time (min)}}
\colnumbers
\startdata
\multicolumn{9}{c}{Band 6 Observations} \\
\hline
\dataset[2013.1.00114.S]{https://almascience.nrao.edu/aq/?project\_code=2013.1.00114.S}&K. I. \"Oberg&1, 2 &2014 July 19&225.650\textendash242.015&32&34\textendash650&43&Continuum\\
\dataset[2013.1.00387.S]{https://almascience.nrao.edu/aq/?project\_code=2013.1.00387.S}&S. Guilloteau&3&2015 May 13 &226.617\textendash244.940& 36 & 21\textendash558 &46& Continuum\\
\dataset[2015.A.00005.S]{https://almascience.nrao.edu/aq/?project\_code=2015.A.00005.S}&T. Tsukagoshi&2&2015 Dec. 1&223.007\textendash242.992&35&17\textendash10804&39& Continuum\\
\hline
\multicolumn{9}{c}{Band 7 Observations} \\
\hline
\dataset[2012.1.00422.S]{https://almascience.nrao.edu/aq/?project\_code=2012.1.00422.S}& E. A. Bergin &4&2015 May 14&330.304\textendash330.539\tablenotemark{a}& 37 & 21\textendash558&20&$^{13}$CO\\
\dataset[2013.1.00196.S]{https://almascience.nrao.edu/aq/?project\_code=2013.1.00196.S}& P. Hily-Blant &-&2014 Dec. 24&330.595\textendash330.653\tablenotemark{a}& 40 & 15\textendash349&38&$^{13}$CO\\
&&&2015 April 04 &330.540\textendash 330.599\tablenotemark{a}&38&15\textendash328&75\\
\dataset[2013.1.00198.S]{https://almascience.nrao.edu/aq/?project\_code=2013.1.00198.S}& E. A. Bergin &5, 6&2014 Dec. 31&337.353\textendash352.011& 34 & 15\textendash349&15& Continuum\\
&&&2015 June 15 &337.299\textendash 352.192&36&21\textendash784&30\\
\dataset[2013.1.01397.S]{https://almascience.nrao.edu/aq/?project\_code=2013.1.01397.S}&D. Ishimoto&7&2015 May 19&329.236\textendash342.904&35&21\textendash539&27& $^{13}$CO + Continuum\\
&&&2015 May 20&329.236\textendash342.904&39&21\textendash 539&27\\
\dataset[2015.1.00686.S]{https://almascience.nrao.edu/aq/?project\_code=2015.1.00686.S}&S. M. Andrews&8&2015 Nov. 23&345.492\textendash358.057&36&23\textendash8259&41& $^{12}$CO + Continuum\\
&&&2015 Nov. 30&345.494\textendash358.059&31&27\textendash10804&43\\
&&&2015 Dec. 1&345.494\textendash358.059&34&17\textendash10804&43\\
\dataset[2016.1.00629.S]{https://almascience.nrao.edu/aq/?project\_code=2016.1.00629.S}& L. I. Cleeves & - & 2016 Dec. 30&345.780\textendash345.839\tablenotemark{a}&45&15\textendash 460&33 & $^{12}$CO\\
&&&2016 Dec. 30 &345.780\textendash345.839\tablenotemark{a}& 45 & 15\textendash 460 & 18\\
&&&2017 July 4 &345.726\textendash345.785\tablenotemark{a}& 44 & 21\textendash 2600 & 44\\
&&&2017 July 9 &345.726\textendash345.785\tablenotemark{a}& 42 & 17\textendash 2600 & 44\\
&&&2017 July 14 &345.727\textendash345.787\tablenotemark{a}& 42 & 19\textendash 1500 & 44\\
&&&2017 July 20 &345.728\textendash345.787\tablenotemark{a}& 42 & 17\textendash 3700 & 44\\
&&&2017 July 21 &345.728\textendash345.787\tablenotemark{a}& 42 & 17\textendash 3700 & 44\\
\enddata
\tablerefs{(1) \citet{2016ApJ...823L..10W}; (2) \citet{2016ApJ...829L..35T}; (3) \citet{2016AA...592A..49T}; (4) \citet{2016ApJ...823...91S}; (5) \citet{2016ApJ...818L..16Z}; (6) \citet{2016ApJ...831..101B}; (7) \citet{2016ApJ...819L...7N}; (8) \citet{2016ApJ...820L..40A}}
\tablenotetext{a}{These observations also included additional spectral windows at other frequencies. The line-free channels were not incorporated into the main continuum analysis in this work due to the relatively small continuum bandwidth available.}
\end{deluxetable*}

\begin{figure*}[htp]
\centering
\includegraphics[scale = 0.65]{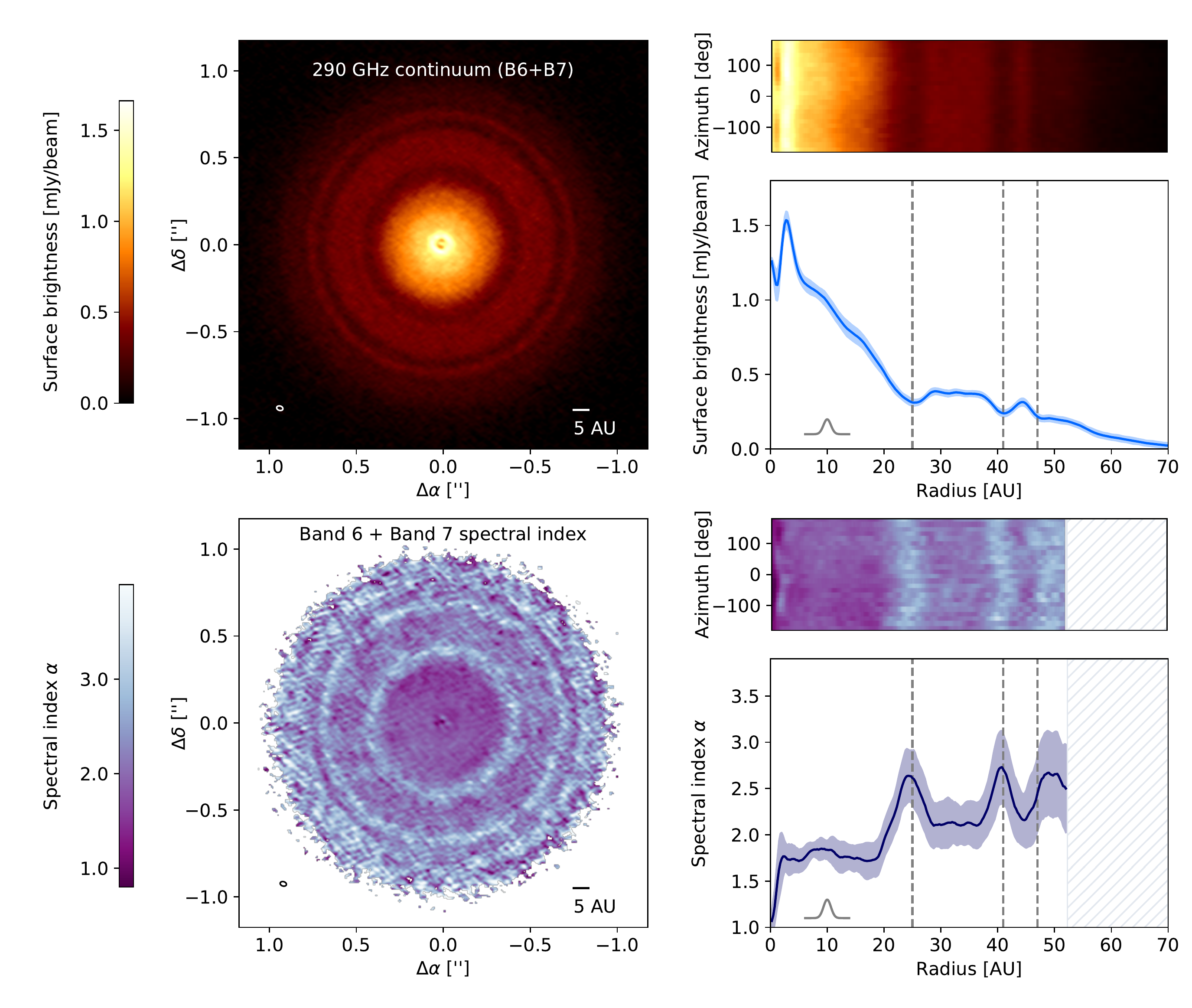}
\caption{\textit{Top left}: 290 GHz continuum image of the TW Hya disk generated from combining Band 6 and Band 7 data. The synthesized beam is shown in the lower left corner. \textit{Top right}: The 290 GHz continuum emission deprojected and replotted as a function of radius and azimuth, with the deprojected, azimuthally averaged radial surface brightness profile shown underneath. The light blue ribbon shows the $1\sigma$ scatter of each radial bin. Gray dashed lines mark the location of the continuum gaps at 25, 41, and 47 AU. The Gaussian profile shows the FWHM of the minor axis of the synthesized beam. \textit{Bottom left}: Spectral index map calculated from Band 6 and Band 7 data. \textit{Bottom right}: The spectral index map deprojected and replotted as a function of radius and azimuth, with its deprojected, azimuthally averaged radial profile shown underneath. The light purple ribbon shows the $1\sigma$ scatter in each radial bin. Values are only shown out to a radius of 55 AU because a substantial fraction of pixels in the continuum image past this radius fall below the signal-to-noise threshold of 5$\sigma$.}
\label{fig:spix}
\end{figure*}

\subsection{$^{12}$CO $J=3-2$ Data Reduction}
The $^{12}$CO $J=3-2$ transition in the TW Hya disk was observed with extended array configurations at channel widths of 244 kHz as part of ALMA program 2015.1.00686.S and with more compact configurations at channel widths of 61 kHz as part of ALMA program 2016.1.00629.S. The raw data for both programs were calibrated by NRAO staff. Details of the observation setups are listed in Table \ref{tab:settings}. 

The phase self-calibration solutions used for the high-resolution 870 $\mu$m continuum imaging were applied to the $^{12}$CO observations from 2015.1.00686.S. The $^{12}$CO observations from  2016.1.00629.S were phase self-calibrated using a continuum model estimated from the line-free channels in the same spectral window. The continuum was then subtracted from the line emission in the visibility plane using the \texttt{uvcontsub} task. Because the two programs were observed with different spectral resolutions, the \texttt{mstransform} task in \texttt{CASA 5.0} was used to regrid and average the visibilities into channels 0.25 km s$^{-1}$ (288 kHz) wide. The $^{12}$CO datasets were then imaged together with the multi-scale CLEAN algorithm (as implemented in the \texttt{tclean} task in \texttt{CASA 5.0}) using scales of 0$''$, 0$\farcs$2, 0$\farcs$4, 0$\farcs$8, 1$\farcs$6, and 3$\farcs$2 and Briggs weighting (robust = 1.0)\footnote{A newer version of \texttt{CASA} was adopted for the line imaging due to a bug in the visibility interpolation of spectral line data in previous versions. See \url{https://casa.nrao.edu/casadocs/casa-5.0.0/introduction/release-notes-50}}. The CLEAN mask was tailored to the emission in individual channels.

The resulting image has a synthesized beam with a FWHM of $139 \times131$ mas ($8.3 \times7.8$ AU) at a position angle of $-74\fdg9$. The rms measured in nearby signal-free channels is $\approx$ 1.7 mJy beam$^{-1}$. A primary beam correction was applied to the image cube with the \texttt{impbcor} task. An integrated intensity map was produced by summing emission above the $3\sigma$ level in the velocity range from $-1.91$ to 7.59 km s$^{-1}$. This velocity range was chosen based on where the emission in the line wings exceeded the 3$\sigma$ level, but the map is robust to choice of integration limits\textemdash truncating or extending the integration range by a few channels changed the integrated flux by less than $0.1\%$.

\section{Observational results}\label{sec:results}
\subsection{The Spectral Index Between 1.3 mm and 870 $\mu$m}
The 290 GHz (Band 6 + Band 7) continuum image, spectral index ($\alpha$) map, and deprojected and azimuthally averaged radial profiles are shown in Figure \ref{fig:spix}.  The adopted position angle and inclination are 152 and 5 degrees, respectively, based on comparisons between spectral line models and data (see Section \ref{sec:models}). These values are consistent within uncertainties with the orientation derived by \citet{2016ApJ...820L..40A} from the 870 $\mu$m continuum. The inclination is slightly smaller than the commonly used value of 7$^\circ$ from \citet{2004ApJ...616L..11Q}, but \citeauthor{2004ApJ...616L..11Q} also use a lower stellar mass value, which is degenerate with inclination. We favor the higher stellar mass value of 0.88 $M_\odot$ derived in \citet{2012ApJ...744..162A} based on SED modeling. Given the low inclination of the TW Hya disk, the deprojected profiles are insensitive to inclination choices within a few degrees of one another.

The continuum emission appears to be azimuthally symmetric and shows prominent gaps at radii of 1, 25, 41, and 47 AU, which is consistent with the Band 6 and Band 7 images published in \citet{2016ApJ...829L..35T} and \citet{2016ApJ...820L..40A}. These and all subsequent measurements quoted from the literature are adjusted for the new \textit{Gaia} distance of 59.5 pc \citep{2016AA...595A...2G}. 

The spectral index profile is mostly flat at radii coinciding with the bright continuum emission rings; $\alpha$ values range from 1.7 to 2.2, taking the scatter into account. The spectral index drops sharply to values close to 1 interior to a radius of 2 AU, which is an unphysical consequence of the innermost gap only being partly resolved in Band 6. The spectral index profile also has local maxima with $\alpha$ peaking at $\sim2.6-2.7$ near the locations of the continuum emission gaps at 25, 41, and 47 AU. \citet{2016ApJ...829L..35T} measured the spectral index of TW Hya at 190 GHz and also reported a steep increase in $\alpha$ at 25 AU, coincident with the most prominent gap in continuum emission. They tentatively suggested enhancements in $\alpha$ at the 41 and 47 AU continuum gaps. Our 290 GHz spectral index map, which has an angular resolution about a factor of two higher than the map from \citet{2016ApJ...829L..35T}, now confirms the presence of the latter two peaks. 

While the spectral index radial profiles around 290 GHz and 190 GHz have similar values in the inner 15 AU of the disk, the $\alpha$ value around 290 GHz is consistently lower than that measured for 190 GHz at a given radius in the outer disk. This behavior is expected based on the dust optical depth increasing with frequency. A more striking difference between the 290 GHz spectral index radial profile presented in this work and that at 190 GHz presented in \citet{2016ApJ...829L..35T} is that the former displays more abrupt changes in $\alpha$ with radius, likely because the 290 GHz map is better resolved. We note that the relative shape of the $\alpha$ radial profile at a given frequency can be determined more robustly than the absolute value of $\alpha$, since the systematic flux calibration uncertainty in each band contributes a constant offset to the entire profile. A $\sim10\%$ flux uncertainty in each band would correspond to an uncertainty in the overall offset of $\Delta \alpha\sim0.4$, but the uncertainty in the relative differences in $\alpha$ is governed by the scatter shown in Figure \ref{fig:spix}.

\begin{figure*}[htp]
\centering
\includegraphics[scale = 0.9]{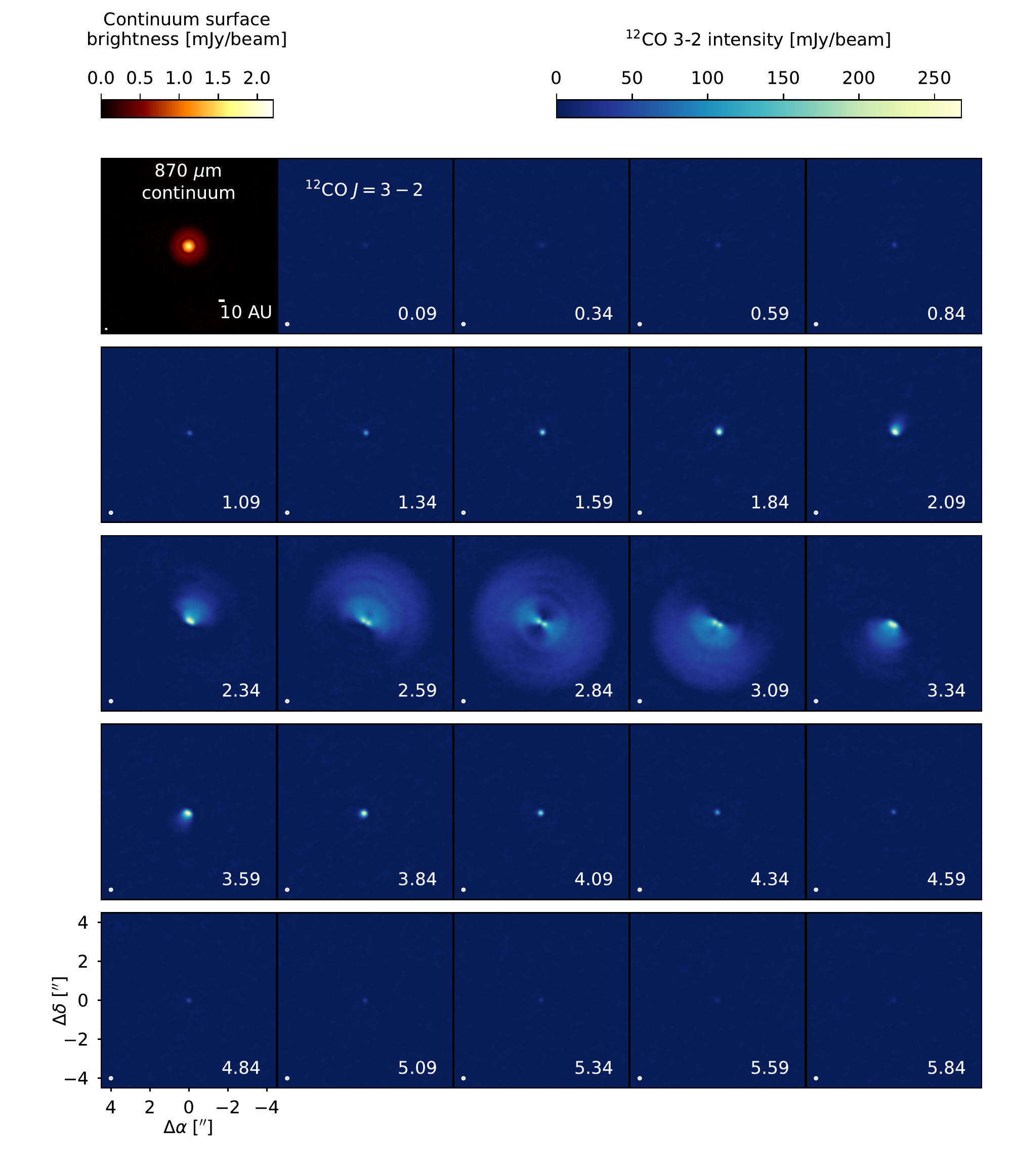}
\caption{Channel maps of the $^{12}$CO $J=3-2$ transition, along with the 870 $\mu$m continuum emission in the upper leftmost panel shown on the same spatial scale. Synthesized beams are drawn in the lower left corner of each panel. The LSR velocity (km s$^{-1}$) is shown in the lower right corner for each $^{12}$CO channel.}
\label{fig:chanmap}
\end{figure*}
\subsection{$^{12}$CO $J=3-2$ Observations}\label{subsec:12COdescription}
\subsubsection{$^{12}$CO $J=3-2$ Emission Morphology}
\begin{figure*}[htp]
\centering
\includegraphics[scale = 0.9]{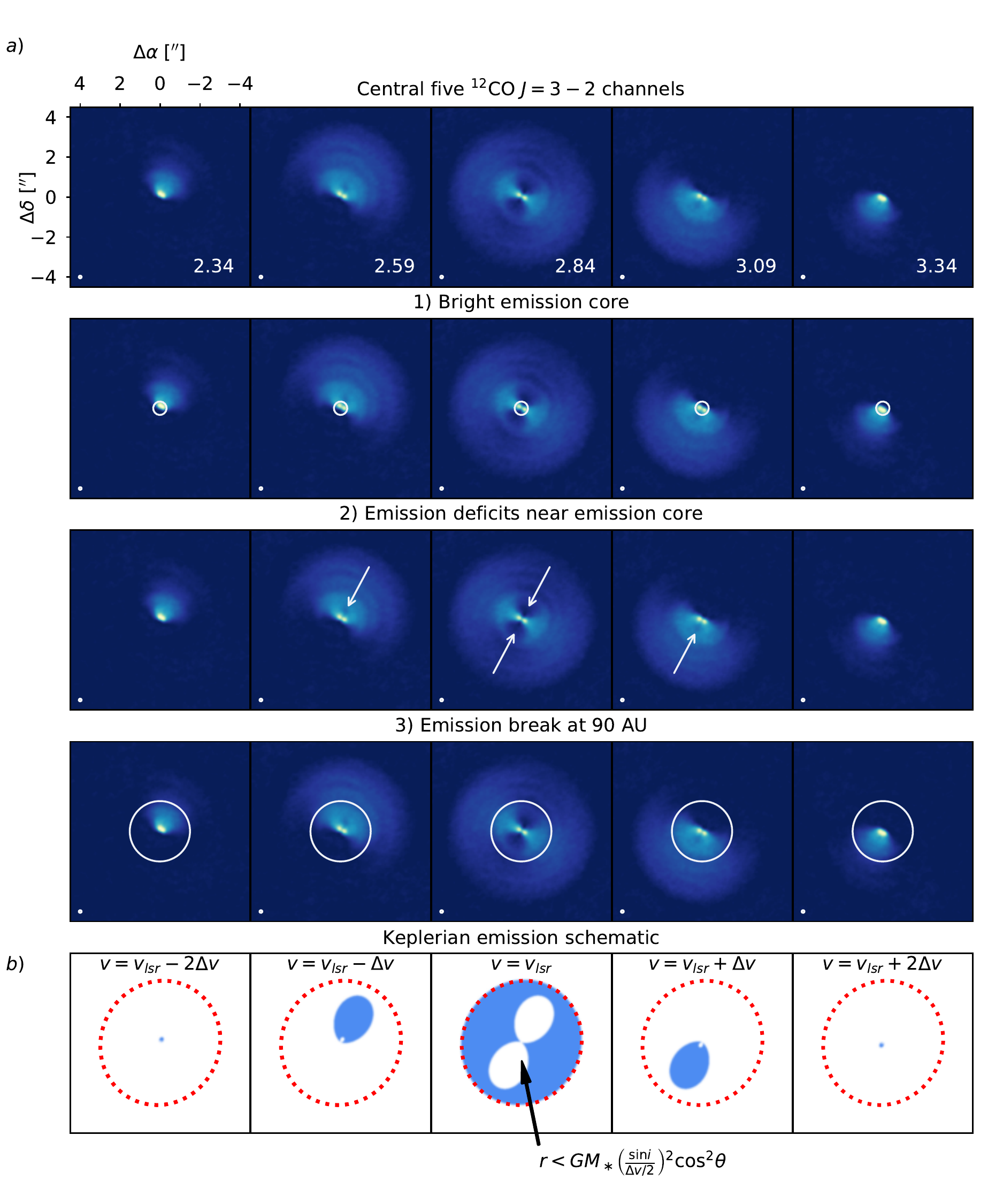}
\caption{\textit{a)} Annotated maps of the central five channels of $^{12}$CO $J=3-2$ to highlight the key emission features (see Figure \ref{fig:chanmap} for context). \textit{b)} Schematic of the line emission morphology of a Keplerian disk (neglecting thermal and turbulent line broadening and beam smearing). The velocity is listed at the top of each panel. The dotted red ellipse marks the boundary of the projected disk. In the central panel at the systemic velocity, the region of the disk that contributes no emission is marked.}
\label{fig:schematic}
\end{figure*}

The $^{12}$CO channel maps are shown in Figure  \ref{fig:chanmap}. $^{12}$CO emission exceeding the $3\sigma$ level extends to a radius of $\approx 3\farcs6$ (215 AU), which is consistent with Submillimeter Array observations reported in \citet{2012ApJ...744..162A}. The CO emission stretches well beyond the submillimeter continuum emission. The channel maps show three key features (see also annotations in Figure \ref{fig:schematic}): 
\begin{enumerate}
\item A bright core of emission extending out to a radius of $\approx25$ AU. 
\item Emission deficits along the disk major axis and adjacent to the bright emission core in the three central channels at 2.59, 2.84, and 3.09 km s$^{-1}$.
\item A break in emission at a radius of $\approx 90$ AU. 
\end{enumerate} 

An additional faint dark arc is visible in several channels at a radius of $\sim125$ AU, but it is not clear whether this is simply an imaging artifact (see Section \ref{sec:fiducial}).   

The emission deficits in the central channels are partly due to the disk inclination (see \citealt{2006AA...448L...5G} and \citealt{2013ApJ...774...16R} for a more detailed discussion). For a geometrically thin disk undergoing Keplerian rotation and being viewed at an inclination angle of $i$, the projected velocity at a point $(r,\theta)$ in the disk is   
\begin{equation}
v_\text{proj}(r,\theta) = v_\text{LSR}+\sqrt{\frac{GM_\ast}{r}}\sin i \cos \theta. 
\end{equation}
Here, $r$ and $\theta$ are in cylindrical coordinates in the frame of the disk, with $\theta=0$ located along the major axis of the disk as seen by an observer. Assuming infinite spatial and spectral resolution and neglecting other line broadening effects, only the region of the disk satisfying 
\begin{equation}\label{eq:emittingregion}
r = GM_\ast \left(\frac{\sin i}{v_\text{proj}-v_\text{LSR}}\right)^2\cos^2\theta.
\end{equation}
contributes to emission observed in a channel at $v_\text{proj}$. 

At the systemic velocity of 2.84 km s$^{-1}$, Equation \ref{eq:emittingregion} collapses into a line along the minor axis of the projected disk, which is aligned with the bright ``hourglass" emission core observed in the TW Hya channel maps. Every channel in reality has non-zero spectral resolution, so it incorporates emission from $v_\text{proj}\pm\frac{\Delta v}{2}$, where $\Delta v$ is the channel width (still neglecting other line broadening effects). Then, the channel at the systemic velocity includes emission from disk regions satisfying 
\begin{equation}
r \geq GM_\ast \left(\frac{\sin i}{\frac{\Delta v}{2}}\right)^2\cos^2\theta.
\end{equation}
This geometry creates emission deficits along the major axis of the disk, near the disk center (see part b of Figure \ref{fig:schematic}), but in Section \ref{sec:models}, we will use radiative transfer models to argue that viewing angle does not wholly account for these features in the TW Hya disk. 

The $^{12}$CO integrated intensity map and corresponding deprojected and azimuthally averaged radial profile are shown in Figure \ref{fig:mommap}. The integrated intensity profile is centrally peaked, but rapidly decreases out to a radius of $\approx$ 30 AU, where the slope of the intensity profile abruptly flattens. The profile exhibits a shoulder at a radius of 70 AU, decreases more rapidly out to a radius of 90 AU, then flattens out and tapers off at 215 AU. The integrated flux, measured inside a circular mask with a radius of $4''$, is 42.7$\pm0.2$ Jy km s$^{-1}$. (The integrated flux uncertainty is estimated with $\sqrt{\text{Area of mask}/\text{Area of beam}}\times \sigma$, where $\sigma=3.8\times 10^{-3}$ Jy km s$^{-1}$ is the rms of the unclipped integrated intensity map.) Taking systematic flux uncertainties into account ($\approx 10\%$), the integrated flux is consistent with previous ALMA and SMA measurements of $^{12}$CO $J=3-2$ in the TW Hya disk \citep{2012ApJ...744..162A, 2012ApJ...757..129R}. The shortest projected baseline is 13.1 m, which corresponds to an angular scale of $\approx14''$ and should therefore adequately recover the large-scale structure in the $^{12}$CO emission. 

We also imaged the line data without continuum subtraction to verify that the key emission features are not artifacts from continuum subtraction, which can create the appearance of line emission substructure if the line optical depth is high enough such that the total outgoing emission has little to no contribution from large dust grains settled near the midplane \citep[e.g.][]{2016PhRvL.117y1101I, 2017ApJ...840...60B}. This effect is not to be confused with optically thick dust absorbing line emission and thereby creating the appearance of molecular emission gaps. In the TW Hya disk, the $^{12}$CO line intensity is substantially larger than that of the dust over many channels, so continuum subtraction does not have a large impact on the observed line emission morphology.

\subsubsection{$^{12}$CO $J= 3-2$ Peak Brightness Temperatures}
\begin{figure*}[htp]
\centering
\includegraphics[scale = 0.65]{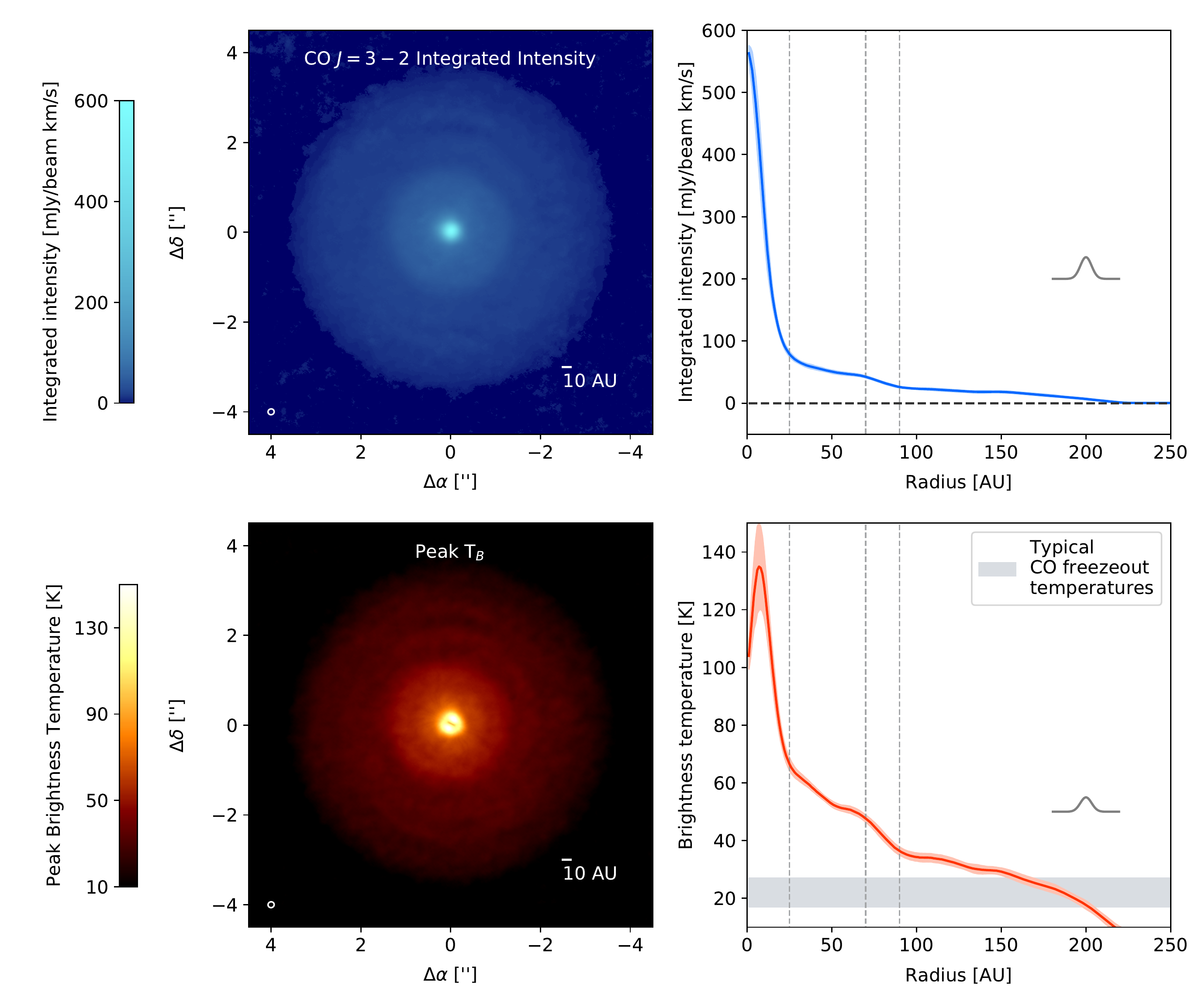}
\caption{\textit{Top left}: An integrated intensity map of the $^{12}$CO $J=3-2$ line. The synthesized beam is shown in the lower left corner.  \textit{Top right}: Deprojected and azimuthally-averaged radial profile of the $^{12}$CO $J=3-2$ integrated intensity. The light blue ribbon shows the $1\sigma$ scatter in each radial bin. The Gaussian profile shows the FWHM of the synthesized beam. \textit{Bottom left}: A peak brightness temperature map of $^{12}$CO $J=3-2$. \textit{Bottom right}: Deprojected and azimuthally-averaged radial profile of the $^{12}$CO $J=3-2$ peak brightness temperature map. The light orange ribbon shows the $1\sigma$ scatter in each radial bin. Vertical gray dashed lines mark the radial break in emission at 25 AU, the shoulder at 70 AU, and the second break at 90 AU. Temperatures between 17 and 27 K are shaded gray to show CO freezeout temperatures that have been estimated for TW Hya (e.g. \citealt{2013Sci...341..630Q} and \citealt{2017NatAs...1E.130Z}).}
\label{fig:mommap}
\end{figure*}

The $^{12}$CO peak brightness temperature map and its deprojected and azimuthally averaged radial profile are also shown in Figure \ref{fig:mommap}. A peak brightness map is computed by taking the maximum value along the frequency axis for each pixel in the image cube and then converting to a brightness temperature using Planck's law (rather than the Rayleigh-Jeans law, which is a poor approximation at these frequencies). Equivalently, 
\begin{equation}
T_\text{B,peak}(x,y) = \max(T_\text{B}(\nu, x,y)),
\end{equation}
where $x$ and $y$ are spatial coordinates and $\nu$ is the channel frequency. In practice, a peak intensity map produced from a single image cube will have minor artifacts tracing the emission boundaries in individual channels. To mitigate this issue, we follow the example of \citet{2014ApJ...785L..12C} and produce three image cubes with the same channel width (0.25 km s$^{-1}$), but with the starting velocities offset by 0.1 km s$^{-1}$. Each image cube yields a peak brightness temperature map. Since the channelization artifacts are spatially offset from one another from map to map, a map with suppressed artifacts can be produced by taking the median value of each pixel from the individual maps.

A peak brightness temperature map for an optically thick line such as $^{12}$CO $J=3-2$ provides an estimate of the gas temperature at the location where the line becomes optically thick, provided that the emission fills the beam. That assumption is valid for most of the disk, which is well-resolved; the exception is the inner 10 AU, where the peak brightness temperature dips. This feature is a consequence of an inclined Keplerian disk being observed at finite angular resolution. Because of the fast rotation of the inner disk, its emission is spread over a large number of channels, with only two narrow wedges contributing to the emission in any given channel (see also \citet{2016ApJ...823...91S} for a related discussion). Referring again to the Keplerian emission schematic in Fig. \ref{fig:schematic}, we see that any Gaussian beam placed over the disk center in a given channel would include regions of the disk contributing no emission.  

The radial profile of the peak brightness temperature map shows an abrupt slope change at a radius of 25 AU, a shoulder at 70 AU, and another abrupt slope change at 90 AU, corresponding to the features observed directly in the channel maps and integrated intensity map. The peak brightness temperatures between radii of 25 and 90 AU range from $\sim$70 to 35 K, which are about twice the value of the gas temperatures derived by \citet{2016ApJ...823...91S} from multiple transitions of the rarer $^{13}$CO and C$^{18}$O isotopologues. \citet{2017NatAs...1E.130Z} estimate that the C$^{18}$O $J=3-2$ flux contribution largely originates from within two gas scale heights of the midplane, in contrast to three to four scale heights for $^{12}$CO $J=3-2$. The difference in inferred gas temperatures for the isotopologues implies a steep vertical temperature gradient.

\subsubsection{Comparison to other CO isotopologues} 
Since CO isotopologues become optically thick at different heights within the disk, they lend insight into the vertical structure.  ALMA observations of the C$^{18}$O and $^{13}$CO $J = 3-2$ transitions in the TW Hya disk have been presented in \citet{2016ApJ...823...91S}, \citet{2016ApJ...819L...7N}, and \citet{2017NatAs...1E.130Z} at spatial resolutions ranging from 0$\farcs$3 to 0$\farcs$5. Although these resolutions are coarser than that of the $^{12}$CO data, they are sufficient to reveal substructure. To facilitate comparisons, we reprocessed and combined archival ALMA observations of $^{13}$CO $J=3-2$ in the TW Hya disk. The reduction details and channel maps are provided in Appendix C. Whereas the integrated intensity maps were presented in \citet{2016ApJ...823...91S} and \citet{2016ApJ...819L...7N}, we now show the channel maps because they more clearly display the weak extended emission features. 

The $^{13}$CO and C$^{18}$O data in \citet{2016ApJ...823...91S}, \citet{2016ApJ...819L...7N}, and \citet{2017NatAs...1E.130Z} show a bright core of emission, an annular gap at $\approx40$ AU, and an outer emission ring peaking at $\approx65$ AU. In the channel maps (see Figure \ref{fig:13COchanmap} for $^{13}$CO and \citet{2017NatAs...1E.130Z} for C$^{18}$O), the annular gap creates central channel emission deficits that coincide spatially with the emission deficits noted for $^{12}$CO. The deficits are more pronounced in $^{13}$CO and C$^{18}$O because they are less optically thick than $^{12}$CO; a similar effect is observed in transition disks, where $^{13}$CO and C$^{18}$O integrated intensity maps have much more prominent central cavities compared to $^{12}$CO \citep[e.g.][]{2015AA...579A.106V, 2016AA...585A..58V}. 

Like the $^{12}$CO channel maps, the $^{13}$CO channel maps in Figure \ref{fig:13COchanmap} also show a steep dropoff in intensity at a radius of $\approx 90$ AU, followed by very faint emission extending out to $\approx200$ AU, which is slightly less extended than the $^{12}$CO emission. C$^{18}$O exhibits an outer emission ring that sharply drops off within a radius of 100 AU \citep{2017NatAs...1E.130Z}, but no emission is observed beyond this radius. The apparent differences between the isotopologues is likely primarily due to sensitivity limits, but may also be partially due to selective photodissociation in the more tenuous outer disk \citep[e.g.][]{2014AA...572A..96M}. 

Based on integrated flux ratios, $^{12}$CO and $^{13}$CO $J = 3-2$ are known to have high optical depths in the TW Hya disk \citep{2013Sci...341..630Q}. However, it is also instructive to determine whether $^{12}$CO is still optically thick beyond a radius of 100 AU, where the emission becomes comparatively weak. As a crude check, we measure the $^{12}$CO/$^{13}$CO integrated flux ratio using an annulus with an inner radius of $1\farcs7$ (101 AU) and an outer radius of $3\farcs6$ (214 AU). Inside this annulus, the $^{12}$CO/$^{13}$CO flux ratio is $\approx20$, which is smaller than the interstellar $^{12}$CO/$^{13}$CO abundance ratio of $\approx 69$ \citep[e.g.][]{1999RPPh...62..143W} and suggests that $^{12}$CO remains optically thick in the outer disk. The $^{12}$CO/$^{13}$CO ratio in the outer disk may be even higher than the interstellar medium (ISM) ratio due to less effective $^{13}$CO self-shielding \citep[e.g.][]{2009AA...503..323V, 2014AA...572A..96M}, in which case the flux ratio still implies that $^{12}$CO is optically thick. 

\begin{figure*}[htp]
\centering
\includegraphics[scale = 0.6]{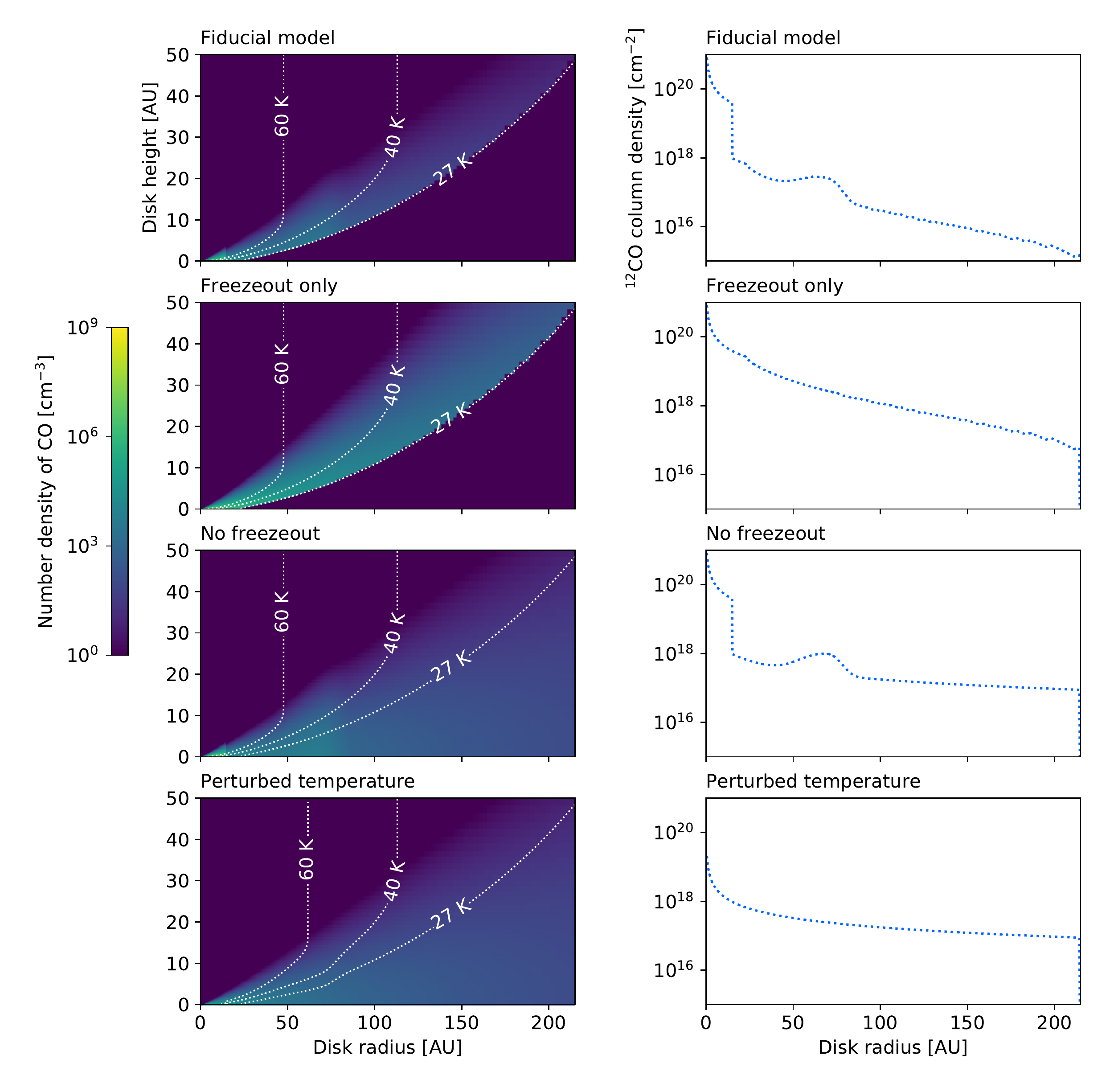}
\caption{Plots of the four parametric model structures used as radiative transfer inputs. \textit{Left}: $^{12}$CO number density color maps with isotherms overplotted as dashed white lines. \textit{Right}: Corresponding $^{12}$CO gas column densities for each model. }
\label{fig:structures}
\end{figure*} 

\section{$^{12}$CO Radiative Transfer Modeling}\label{sec:models}

\subsection{Overview}

To explore possible origins for the $^{12}$CO emission substructure, we perform radiative transfer calculations based on several sets of parametric structure models. Given the complex details visible at high resolution, as well as the large computational cost of synthesizing the image cubes, our aim in this work is not to present an optimal model, but to provide guidance for which physical and chemical mechanisms would plausibly yield the key emission features noted in Section \ref{sec:results}. Similar modeling approaches have been favored for ALMA observations of molecular line emission in a number of works \citep[e.g.][]{2012ApJ...757..129R, 2013ApJ...774...16R,2016PhRvL.117y1101I, 2017ApJ...839...43O} because of the flexibility offered in directly specifying temperature structures and molecular abundances to explore particular emission features of interest. An alternative approach is thermochemical disk modeling, which computes dust and gas structures and molecular abundances in a physically self-consistent manner via numerical codes such as ProDiMo \citep{2009AA...501..383W} or DALI \citep{2012AA...541A..91B}, although such an approach is also subject to uncertainties in the values adopted for parameters such as dust opacities or reaction rates \citep[e.g.][]{2008ApJ...672..629V, 2017AA...607A..41K}. The parametric and thermochemical approaches are ultimately complementary; physical models guide the setup of parametric models, the results of which then motivate additional physical modeling \citep[e.g.][]{2012ApJ...744..162A, 2013Sci...341..630Q, 2014ApJ...780..153B,2017AA...599A.101V}.

Because continuum subtraction did not appear to introduce artifacts to the $^{12}$CO image cubes, we model the continuum-subtracted line emission rather than line+continuum, which would require a number of additional assumptions and free parameters to model the dust. As discussed in \citet{2017ApJ...840...60B}, continuum subtraction tends to remove a larger fraction of $^{13}$CO and C$^{18}$O line emission because these lines are less bright overall but are still optically thick in the inner disk. Since a detailed physical model for the continuum would likely be necessary to construct a CO model consistent with all isotopologue data, we focus on modeling $^{12}$CO $J=3-2$ only in the present paper, and defer dust modeling and multi-line fitting to future work. 

Perhaps the most straightforward explanation of the $^{12}$CO emission features is that they trace an  initial sharp drop in the CO column density, which then rebounds in the outer disk. Such a column density profile has been inferred previously based on observations of less optically thick CO isotopologues in the TW Hya disk \citep{2016ApJ...819L...7N, 2016ApJ...823...91S, 2017NatAs...1E.130Z}. Hence, our fiducial $^{12}$CO model is motivated by the C$^{18}$O $J=3-2$ model presented in \citet{2017NatAs...1E.130Z}. We also construct a model with a monotonically decreasing CO column density profile to demonstrate how its emission features differ from the fiducial model. We then examine the impact of our assumptions about CO freezeout. Finally, we briefly consider how assumptions about the form of the temperature structure affect inferences about the CO column density profile.

We make several assumptions common to all models. Because there are no clear azimuthal asymmetries in the ALMA data, all the models are axisymmetric and specified in cylindrical coordinates. After some experimentation, the inclination and position angle were fixed to 5 and 152 degrees, respectively.  Local thermal equilibrium is assumed for calculating the intensity of $^{12}$CO $J = 3-2$ because its critical density ($\sim10^4$ cm$^{-3}$) is small relative to typical disk gas densities \citep[e.g.][]{2007ApJ...669.1262P}. 

To first order, the gas velocity field is that of a thin Keplerian disk in which the stellar mass greatly exceeds the disk mass:
\begin{equation}
\begin{cases}
& v_r = v_z = 0\\
& v_\phi = \sqrt{\frac{GM_\ast}{r}}.
\end{cases}
\end{equation}

As in \citet{2017NatAs...1E.130Z}, we fix the microturbulent line broadening parameter to $a_\text{turb}$ = 0.01 km s$^{-1}$, but we note that this value was motivated by general disk theory rather than from a direct measurement, and in practice this parameter likely varies spatially. The value of this parameter is not well-established\textendash estimates for the TW Hya disk have been disparate due to differences in methodology as well as limitations in instrumental resolution and precision \citep{2011ApJ...727...85H, 2016AA...592A..49T}. The turbulence values that have been estimated for the TW Hya disk are smaller than the spectral resolution of our observations, limiting their utility in constraining turbulence independently. In terms of modeling line emission, turbulence is highly degenerate with the thermal structure (see also \citealt{2015ApJ...808..180S} for a more detailed exploration). Holding all other parameters equal, increasing the turbulent broadening parameter by a factor of 10 (bringing it up to the higher values estimated by \citealt{2016AA...592A..49T}) makes the emission features ``blurrier'' and ``fills in'' the central emission deficits in the channel maps, but the difference is modest and does not qualitatively change our interpretation of the CO observations.

The largest projected velocities where $^{12}$CO emission is detected ($\pm4.75$ km s$^{-1}$ from the systemic velocity) suggest that there is gas within at least a few tenths of an AU from the central star, assuming $v_\text{proj} = \sqrt{\frac{GM_\ast}{r}}\sin(i)$. In the absence of more precise information about the gas inner radius, we set $R_\text{in}$ to 0.05 AU, roughly the inferred location of the inner dust rim \citep{2006ApJ...637L.133E}. The outer radius $R_\text{out}$ is set to 215 AU, based on the extent of $^{12}$CO emission above the $3\sigma$ level in the channel maps.  

Between the inner and outer radii, we assume that the vertical distribution of the gas is approximated by a Gaussian with a standard deviation $H_\text{mid}(r)$, where

\begin{equation}
H_\text{mid}(r) = \sqrt{\frac{k_B T_\text{mid}(r)r^3}{\mu_\text{gas}m_\text{H} G M_\ast}}
\end{equation}
is the gas pressure scale height, $T_\text{mid}(r)$ is the midplane temperature, $\mu_\text{gas}=2.37$ is the mean mass for a gas particle, and $m_\text{H}$ is the mass of atomic hydrogen. We therefore express the $^{12}$CO number density as 
\footnotesize
\begin{equation}\label{eq:numdens}
n_\text{CO}(r,z) = p(T(r,z))\times \frac{N(r)}{\sqrt{2\pi} H_\text{mid}(r)}\exp\left[-0.5\left(\frac{ z} {H_\text{mid}(r) }\right)^2\right].
\end{equation} 
\normalsize
$N(r)$ is a scaling factor that simplifies to the $^{12}$CO column density profile in the case where $p(T)=1$ everywhere (e.g., in the absence of freezeout). The step function $p(T)$ accounts for CO freezeout, such that $p=1$ for gas temperatures above some freezeout temperature $T_\text{frz}$ and $p = 10^{-6}$ everywhere else. This procedure, similar to the treatment in \citet{2008ApJ...681.1396Q}, divides the disk into a warm molecular layer with abundant CO and a cold midplane layer where freezeout depletes gas-phase CO. In general, UV photodissociation sets the upper boundary of the CO distribution and plays a role in setting the outer boundary \citep[e.g.][]{1999AA...351..233A}. Because ambiguities in the gas and grain size distribution (see Section \ref{sec:discussion}) render explicit calculations of the photodissociation boundary difficult, we elect to parameterize the CO distribution directly.

The radiative transfer code RADMC-3D\footnote{\url{http://www.ita.uni-heidelberg.de/~dullemond/software/radmc-3d/}} \citep{2012ascl.soft02015D} is used to compute the $^{12}$CO level populations and perform the raytracing to produce model image cubes. The parametric expressions for the temperature, gas velocity, and $^{12}$CO number density are evaluated at 400 logarithmically spaced radial bins from 0.05 to 400 AU and at 100 logarithmically spaced polar angle bins from 0 to $\pi/2$ in a spherical coordinate system (note that the grid points are converted to cylindrical coordinates before the expressions are evaluated, and mirror symmetry is assumed for the upper and lower halves of the disk). The molecular data inputs for the radiative transfer are obtained from the LAMDA database \citep{2005AA...432..369S}. To account for the effects of non-zero channel widths on the observed spatial distribution of the line emission, the model image cubes are synthesized at a velocity resolution of 0.05 km s$^{-1}$ and subsequently averaged to a resolution of 0.25 km s$^{-1}$ to match the observations.  The \texttt{vis\_sample} package\footnote{Version used in this work available at \url{https://github.com/AstroChem/vis_sample/tree/j}. General version available at \url{https://github.com/AstroChem/vis_sample}.} \citep{2015ApJ...806..154C, loomis} is used to sample the radiative transfer images at the same spatial frequencies as the data in order to produce model visibilities, which are then imaged with the same procedure described in Section 2. 

 \begin{deluxetable}{ccccc}
\tabletypesize{\scriptsize}
\tablecaption{Parameter Values for $^{12}$CO Models\label{tab:models}}
\tablehead{\colhead{Parameter}&\colhead{Fiducial}&\colhead{Freezeout}&\colhead{No}&\colhead{Perturbed}\\
&&\colhead{only}&\colhead{freezeout}&\colhead{temperature}}
\startdata
$R_\text{in}$ (AU) & 0.05& 0.05& 0.05& 0.05\\
$R_\text{out}$ (AU) & 215& 215& 215& 215\\
$\gamma$ &0.9&0.9&0.9&0.9\\
$N_\text{CO}$ (cm$^{-2}$)&$3\times 10^{19}$&$3\times 10^{19}$&$3\times 10^{19}$&$7.5\times 10^{17}$\\
$R_1$ (AU) & 15 & - & 15 & 15\\
$R_2$ (AU) & 70 & - & 70 & 70\\
$A$ & 3 & - & 3 & -\\
$f$ & 0.025 & - & 0.025& -\\
$\sigma_\text{in}$&12&- & 12 & -\\
$\sigma_\text{out}$&6&-& 6 & -\\
$T_{\textup{mid},10}$ (K)&40&40 & 40 & 40 \\
$T_{\textup{atm},10}$ (K) &125&125 & 125 & 125\\
$q$& 0.47 &0.47 & 0.47 & 0.47\\
$\delta T$ & - & - & - & 1.6\\
$B$ & - & - & - & 0.15\\
$\sigma_{T_\text{in}}$&-&-&-&15\\
$\sigma_{T_\text{out}}$&-&-&-&6\\
$T_\text{frz}$ (K) & 27 & 27 & - & -\\
$a_\text{turb}$ (km s$^{-1}$) & 0.01 & 0.01 & 0.01 & 0.01 
\enddata
\end{deluxetable}

\begin{figure*}[htp]
\begin{center}
\includegraphics[scale = 0.57]{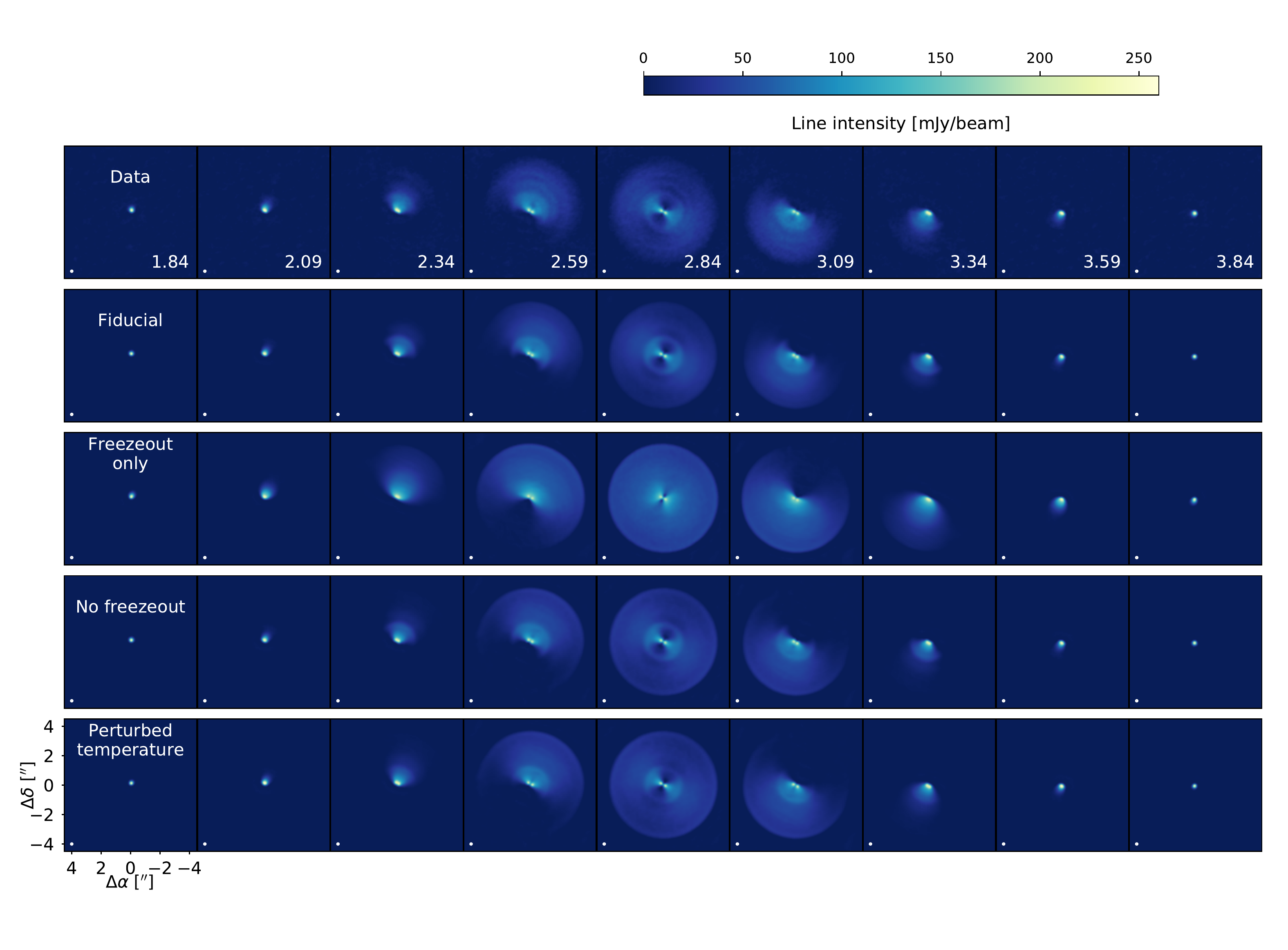}
\end{center}

\caption{Central channels of the $^{12}$CO $J=3-2$ observations, compared with model channel maps. Synthesized beams are drawn in the lower left corner of each panel. The LSR velocity (km s$^{-1}$) is shown in the lower right corner of each panel in the first row. The offset from phase center is marked in arcseconds in the lower left panel. }
\label{fig:modelimages}
\end{figure*}

\begin{figure*}[htp]
\begin{center}
\includegraphics[scale = 0.75]{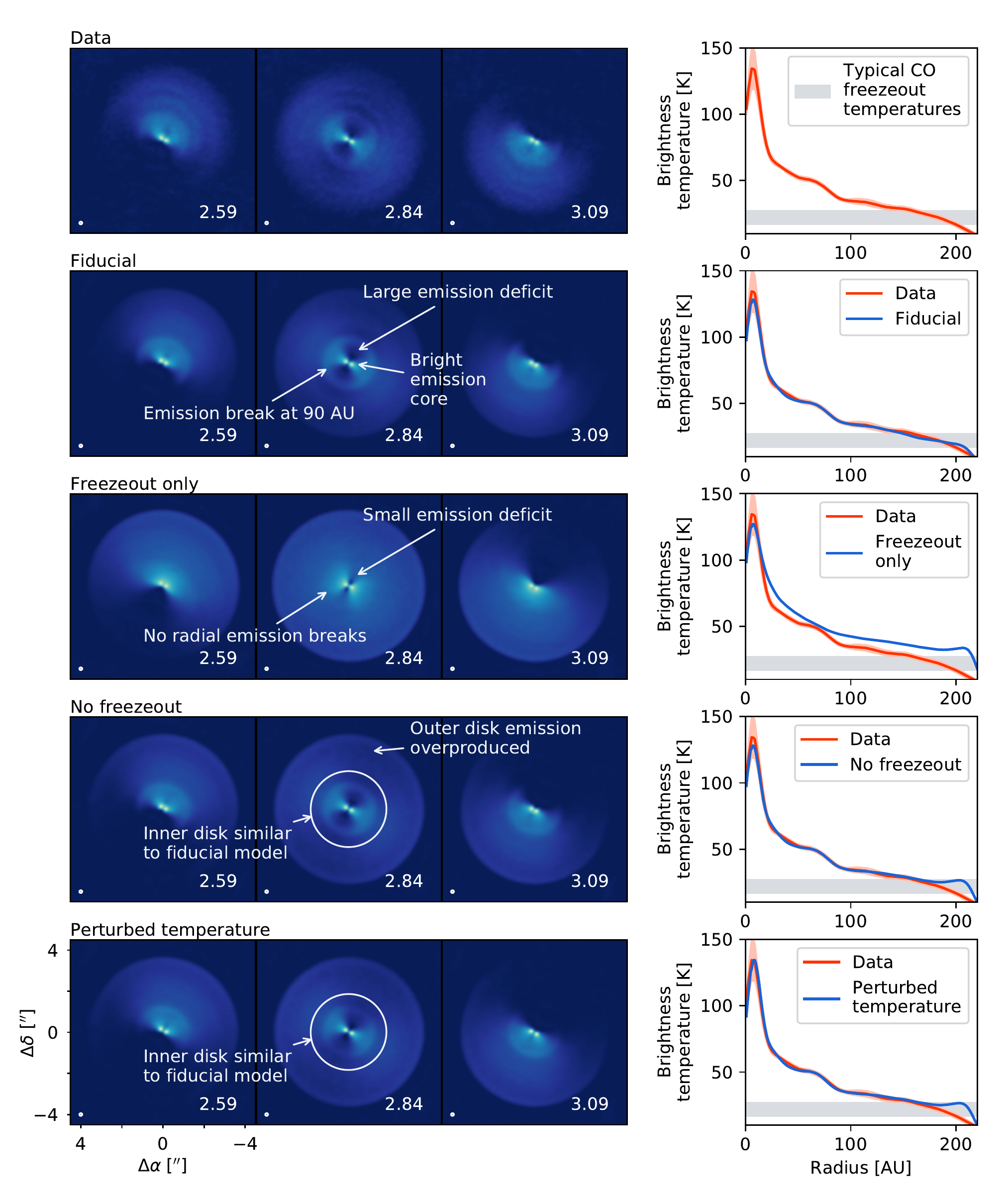}
\end{center}
\caption{Insets of the three central channels for each model to highlight the substructures, along with the corresponding radial profile of the peak brightness temperature map.}
\label{fig:insets}
\end{figure*}

The model parameters are listed in Table \ref{tab:models}, and the gas temperatures and $^{12}$CO number and column densities for each model are plotted in Figure \ref{fig:structures}. Channel maps made from the model visibilities are shown in Figure \ref{fig:modelimages}, beneath the corresponding channels from the observations. To highlight some details of the models, insets of the central three channels and radial profiles of the peak brightness temperature maps are shown in Figure \ref{fig:insets}.

\subsection{The Fiducial Model\label{sec:fiducial}}
The parameterization and values for the fiducial model are motivated by the C$^{18}$O column density profile derived in \citet{2017NatAs...1E.130Z}\textemdash the CO column density decreases sharply in the inner disk, then features a secondary peak in the outer disk. Functionally, this is specified by setting the scaling factor for the number density to be
\begin{equation}\label{eq:scaling}
N(r) = N_\text{CO}\left(\frac{r}{\text{20 AU}}\right)^{-\gamma}\times f_1(r)\times f_2(r).
\end{equation}
$f_1(r)$ is a factor setting the radial location and degree of the column density drop in the inner disk:
\begin{equation}
f_1(r)=  \begin{cases} 
      1, & r< R_1 \\
      f, & r\geq R_1.
   \end{cases}
\end{equation}
$f_2(r)$ sets the shape of the secondary bump in the $^{12}$CO column density profile such that 
\begin{equation}\label{eq:f2}
f_2(r) = \begin{cases} 
      1+A\exp{\left( -\frac{(r-R_2)^2}{2\sigma_\text{in}^2}\right)}, & r< R_2 \\
        1+A\exp{\left( -\frac{(r-R_2)^2}{2\sigma_\text{out}^2}\right)}, & r\geq R_2.  
   \end{cases}
\end{equation}
The asymmetric Gaussian shape of the secondary bump is motivated by the shallow slope of the CO integrated intensity radial profile inward of 70 AU and the much steeper slope outside of 70 AU, although other parameterizations may achieve similar effects. $N_\text{CO}$ is set to $3\times10^{19}$ cm$^{-2}$, assuming that the $^{12}$CO to C$^{18}$O abundance ratio follows the local ISM value of $\sim500$ \citep[e.g.][]{1999RPPh...62..143W}. The model parameter values are listed in Table \ref{tab:models}.

Because LTE is assumed, the absolute $^{12}$CO number density distribution is directly input into RADMC-3D without requiring the underlying gas distribution to be specified. The parameterization for $^{12}$CO in Equations \ref{eq:numdens} and \ref{eq:scaling} can therefore be viewed in two equivalent ways:
\begin{enumerate}
 \item The fractional CO abundance, $X_\text{CO}$, is constant in the warm molecular layer, and perturbations in the gas surface density profile govern the radial variation in the $^{12}$CO column density profile. An ``unperturbed'' surface density profile is assumed to follow a power-law, as outlined in \citet{1974MNRAS.168..603L} for a viscous disk. More colloquially, these perturbations might be referred to as an annular gap and ring in the gas distribution.
 \item The gas surface density profile follows a standard power law, and radially varying CO depletion (i.e. reductions in $X_\text{CO})$ in the warm molecular layer creates the radial variations in the $^{12}$CO column density profile. 
\end{enumerate}

The temperature structure is modeled using the vertical gradient prescription presented in \citet{2003AA...399..773D} and \citet{2012ApJ...744..162A}: 
\footnotesize
\begin{equation}
T(r,z) =  \begin{cases} 
      T_{\textup{atm}}(r)+(T_{\textup{mid}}(r)-T_{\textup{atm}}(r))\cos^4{\left(\frac{\pi z}{2 z_q}\right)} & z\textless z_q \\
     T_{\textup{atm}}(r) & z\geq z_q,
   \end{cases}
\end{equation}
\normalsize
where
\begin{equation}
T_\textup{atm}(r) = T_{\textup{atm},10}\left(\frac{r}{\textup{10 AU}}\right)^{-q},
\end{equation}
\begin{equation}
T_\textup{mid}(r) = T_{\textup{mid},10}\left(\frac{r}{\textup{10 AU}}\right)^{-q},
\end{equation}

and 

\begin{equation}
z_q = 4H_\text{mid}(r).
\end{equation}
\citet{2017NatAs...1E.130Z} employ a vertically isothermal model, assuming that C$^{18}$O emits at the midplane temperature. In contrast, $^{12}$CO becomes optically thick well above the midplane, requiring the adoption of a vertical temperature gradient. We set $T_{\textup{mid},10} = 40$ K and $q = 0.47$ based on the midplane temperature derived in \citet{2017NatAs...1E.130Z}. $T_{\textup{atm},10}$ is less constrained; we choose a value of 125 K such that the spatially integrated fluxes of the channels at the systemic velocity for the observations and fiducial model are within 5$\%$ of one another. This value is broadly in line with the detailed temperature structures computed from fitting the SED \citep[e.g.][]{2015ApJ...799..204C, 2016AA...592A..83K}. The consequences of our assumptions about the temperature structure are discussed later.

The fiducial model images, shown in Figures \ref{fig:modelimages} and \ref{fig:insets}, reasonably reproduce the three key features of the observed channel maps\textemdash  the sharp drop in the $^{12}$CO column density at $r = 15$ AU creates the bright ``hourglass" emission core as well as the nearby emission deficits in the central channels, and the secondary bump in the column density creates the emission shoulder at $r\approx70$ AU and the radial intensity break at $\approx$ 90 AU. Whereas specifying a sharp drop in the CO column density creates a clear annular gap in models of  C$^{18}$O emission \citep{2017NatAs...1E.130Z}, the impact is more subtle in $^{12}$CO emission due to its much higher optical depth. In the optically thin limit, intensity scales almost linearly with column density. In contrast, for optically thick $^{12}$CO, altering the number density changes the height and therefore the temperature of its emitting surface in the disk, which yields emission substructure. It should be noted that the drop in the column density at $r = 15$ AU reproduces the break in the emission profile slope at $r\approx25$ AU because the beam smears out the radial change in intensity.

The model slightly underestimates the line intensity in the inner few AU of the disk. This may be due to complexities in the vertical structure that are unaccounted for in the model, deviations from Keplerian rotation in the inner disk \citep{2012ApJ...757..129R}, or the slightly differential rotation associated with a geometrically thick disk \citep[e.g.][]{2013ApJ...774...16R}. While our simple models are aimed at elucidating features 15 AU and beyond, the effect of these secondary considerations on the inner disk would be interesting to explore in future work.

In the outer disk past a radius of 90 AU, there are additional faint arcs visible in the central three channels even though the surface density profile is smooth at these radii. These appear to be artifacts from imaging the visibilities with CLEAN. Similar faint arcs appear in the outer disk emission in the $^{12}$CO observations and may be imaging artifacts as well. However, since the arcs in the data persist with different choices of imaging parameters and do not exactly match the artifacts in the model images, it is difficult to say for certain whether there is any genuine physical origin for these features.

\subsection{A Freezeout-only Model}
While the fiducial model illustrates that substructure in the CO column density profile is compatible with the observed $^{12}$CO emission, it does not by itself demonstrate that substructures are \textit{necessary} to create the observed emission features. Given the high optical depth of $^{12}$CO $J = 3-2$, it is not intuitive that column density substructure would have a visible impact on its emission morphology. It is therefore instructive to compare the fiducial model with a ``freezeout-only" model, where the $^{12}$CO column density profile decreases monotonically from the disk center and no depletion occurs in the warm molecular layer.  

The $^{12}$CO number density scaling factor is now
\begin{equation}\label{eq:smooth}
N(r) = N_\text{CO}\left(\frac{r}{\text{20 AU}}\right)^{-\gamma}, 
\end{equation}
i.e., it is an extrapolation of the profile in the inner disk of the fiducial model. The ``freezeout only" model is otherwise specified in the same fashion as the fiducial model. This ``freezeout only" model is similar to the parameterization that has been used to fit CO emission in a number of protoplanetary disks observed at coarser angular resolution \citep[e.g.][]{2012ApJ...744..162A, 2014ApJ...788...59W, 2015ApJ...813...99F}. 

The resulting model channel maps, shown in Figures \ref{fig:modelimages} and \ref{fig:insets}, differ dramatically from those of the observations and fiducial model. The failure of the ``freezeout-only" model to reproduce the key features of the observations indicates that these features are not artifacts resulting from sparse $uv$ coverage or of the deconvolution algorithm, but reflect the structure of the disk itself. No intensity break appears at a radius of 90 AU, since the surface density profile is smooth there. While a bright emission core is present due to the high temperatures of the inner disk, the core is not as sharply defined as in either the observations or the fiducial model. Though emission deficits near the center of the disk are visible at the systemic velocity, they are substantially smaller than the ones in the observations and fiducial model. As remarked upon in Section 2, emission deficits at those locations are expected for inclined disks in Keplerian rotation, but the discrepancy between the scale of the deficits for the fiducial and ``freezeout-only" models suggests that disk orientation alone does not explain the observed emission morphology\textemdash an abrupt change in the CO column density and/or temperature profile has to occur near $r\approx15$ AU to simultaneously create the bright inner emission core and adjacent emission deficits. 

\begin{figure}[htp]
\centering
\includegraphics[scale = 0.4]{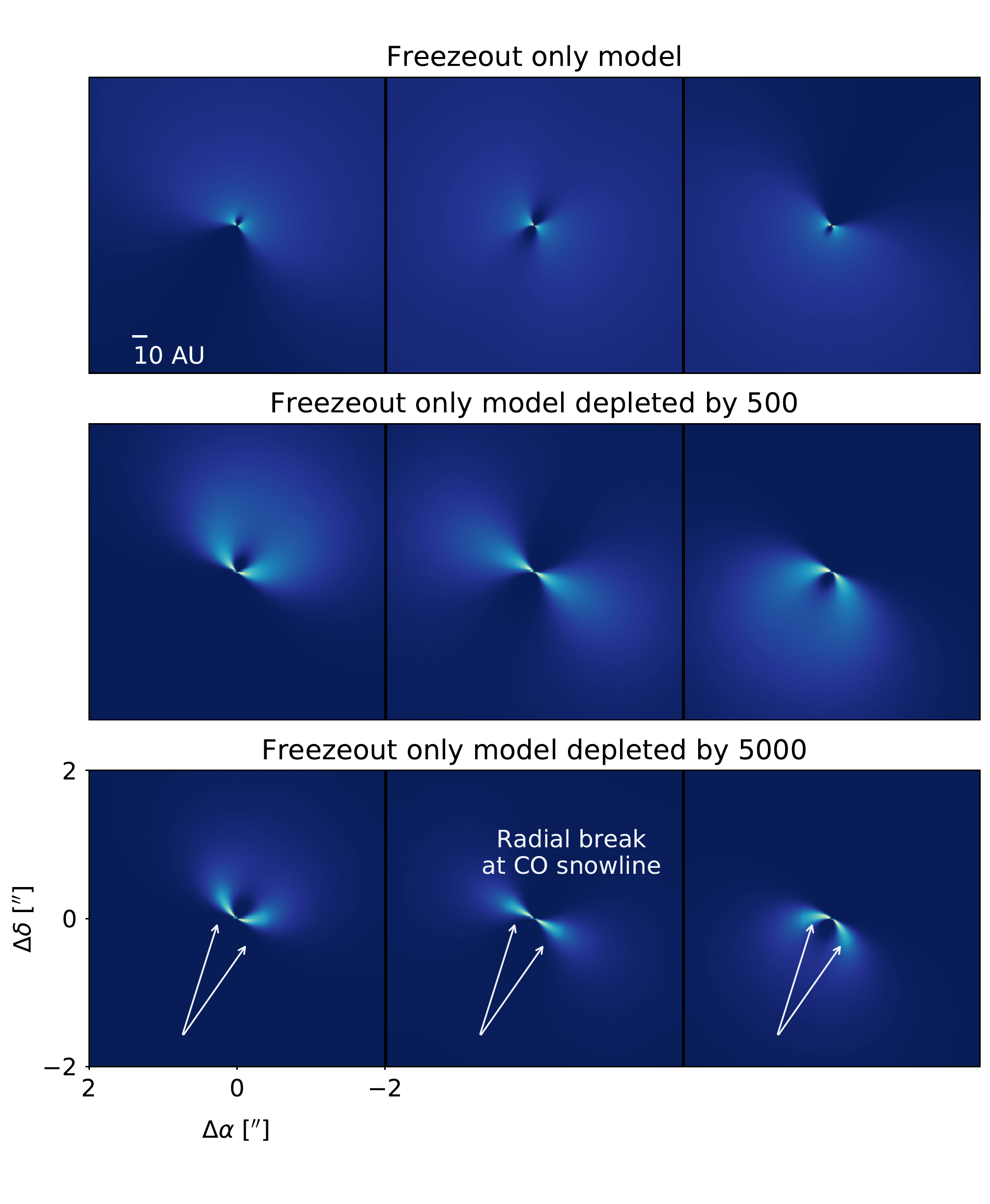}

\caption{Comparison of the central three channels of the RADMC-3D ``freezeout only" images (zoomed in to the inner 2$''$ of the disk) with models where $N_\text{CO}$ is reduced by a factor of 500 and 5000, respectively. Only the model depleted by a factor of 5000 shows a radial intensity break corresponding to the location of the CO snowline.}
\label{fig:depletion}
\end{figure}

For this ``freezeout-only'' model, the midplane CO snowline does not create a marked intensity break because the $^{12}$CO emission is too optically thick. $N_\text{CO}$, which sets the overall scale of the column density profile, has to be reduced by a factor of several thousand for the CO snowline to become visible in the images generated by RADMC-3D, and even then the effect is subtle (see Figure \ref{fig:depletion}). In contrast, the fiducial model only invokes a factor of 40 reduction in the $^{12}$CO column density at a radius of 15 AU to create emission substructure by lowering the CO emitting height. While CO column densities are not well-constrained by $^{12}$CO, the value of $N_\text{CO}$ chosen for the unreduced ``freezeout only'' model is based on column density estimates from observations of C$^{18}$O, which has much lower optical depth \citep{2017NatAs...1E.130Z}. Since it is unlikely that C$^{18}$O column densities have been overestimated by several orders of magnitude for the TW Hya disk, severely reducing $N_\text{CO}$ to decrease the optical depth of $^{12}$CO does not seem to be a viable route for reproducing the observed $^{12}$CO emission morphology.

\subsection{A Model With No Freezeout\label{sec:nofreezeout}}
While the fiducial model adopts a CO freezeout temperature of 27 K to be consistent with the C$^{18}$O model from \citet{2017NatAs...1E.130Z}, \citet{2013Sci...341..630Q} previously advocated for a freezeout temperature of 17 K in the TW Hya disk. In practice, CO desorption occurs at various temperatures over this range, depending on the substrate \cite[e.g.][]{1988Icar...76..201S, 2003ApJ...583.1058C, 2016ApJ...816L..28F}. Hence, it is worthwhile to examine how the model emission depends on the treatment of CO freezeout. To illustrate how an extreme change in the CO freezeout specification could look, we produce a ``no-freezeout" model by using a CO number density structure in which $p(T) = 1$ everywhere in Equation \ref{eq:numdens}. In all other respects, the ``no-freezeout" model is identical to the fiducial model. 

The resulting model channel maps are shown in Figure \ref{fig:modelimages}. Past $r\approx150$ AU, the model images overproduce emission in comparison with the observations and the fiducial model, but the key features of the $^{12}$CO emission (i.e., the bright core, the nearby emission deficits in the central channels, and the break in intensity at 90 AU) look similar. This suggests that the $^{12}$CO emission morphology in the inner disk is not sensitive to differences in assumptions about CO freezeout. 

One could perhaps devise a colder temperature structure that removes so much gas-phase CO that the optical depth of $^{12}$CO is low enough to reveal its midplane snowline. However, this possibility for TW Hya can be ruled out by the $^{12}$CO peak brightness temperature map in Figure \ref{fig:mommap}. Within a radius of 100 AU, the peak brightness temperatures are well above the range of expected CO freezeout temperatures, indicating that the $^{12}$CO emitting surface is at a height much warmer than the snow surface and therefore does not trace the onset of freezeout in the midplane. 

\subsection{A ``Perturbed Temperature" Model}
The models we have presented so far sought to reproduce the observed $^{12}$CO emission morphology by shifting the CO emitting height through surface density substructures. An alternative route is to modify the disk temperature structure directly. The temperature structure used for the previous models has been relatively simple, with the radial temperature dependence following a standard power law for a fixed height. However, models from several works suggest that the TW Hya temperature structure could be much more complex. First, scattered light observations of the TW Hya disk show multiple annular gaps \citep[e.g.][]{2013ApJ...771...45D, 2015ApJ...802L..17A, 2015ApJ...815L..26R, 2017ApJ...837..132V}. Relative to a smooth disk model, the troughs of gaps have lowered temperatures due to shadowing from the interior wall of the gap, while the far walls of gaps receive more stellar radiation and therefore feature elevated temperatures \citep[e.g.][]{2012ApJ...749..153J, 2013ApJ...772...34J}. Additional cooling of the gas may occur within the gap if dust densities decrease to the point where the gas and dust temperatures are no longer coupled via collisions \citep{2017arXiv171004418F}. Second, inward radial drift of larger solids in disks may create radial temperature inversions; \citet{2016ApJ...816L..21C} discusses how radial drift reduces dust optical depths in the outer disk and allows stellar radiation to penetrate deeper in the disk, while \citet{2017AA...605A..16F} suggest that radial drift allows the outer disk to be heated more efficiently because the remaining small grains are lofted upward and receive more stellar illumination. 

To illustrate how thermal and density variations can create similar emission patterns, we construct a model that has a smooth CO column density profile. The bright inner core of emission and the intensity break at 90 AU seen in the data are then reproduced by increasing the temperatures in these regions relative to the fiducial model. The $^{12}$CO number density scaling factor $N(r)$ is again described by a power law, i.e. Equation \ref{eq:smooth}. $N_\text{CO}$ is set to $7.5\times 10^{17}$ cm$^{-2}$ such that the $^{12}$CO column density profile matches that of the disk outside 15 AU in the ``no-freezeout'' model, excluding the secondary column density bump. In order to keep the column density profile smooth, CO freezeout is ignored in this model. As shown in the ``no-freezeout" model, the treatment of CO freezeout does not have a noticeable effect on the inner 150 AU of the disk, which is where the substructure is observed.

The modified temperature structure is 

\begin{equation}
T_\text{perturbed}(r,z) = T_\text{fiducial}\times g_1(r)\times g_2(r).
\end{equation}
The prescription for $g_1(r)$ creates a hot inner disk, similar to a TW Hya model from \citet{2012ApJ...757..129R} : 
\begin{equation}
g_1(r) = \begin{cases}
\delta T & r< R_1 \\
1 & r\geq R_1. 
 \end{cases}
\end{equation}
$g_2(r)$ creates a secondary warm region in the outer disk:
\begin{equation}
g_2(r) =       \begin{cases} 
      1+B\exp{\left( -\frac{(r-R_2)^2}{2\sigma_{T_\text{in}}^2}\right)}, & r< R_2 \\
        1+B\exp{\left( -\frac{(r-R_2)^2}{2\sigma_{T_\text{out}}^2}\right)}, & r\geq R_2.  
   \end{cases}\end{equation}
Parameters $g_1(r)$ and $g_2(r)$ are the temperature analogues to $f_1(r)$ and $f_2(r)$, the factors setting the CO column density drop in the inner disk and secondary bump in the outer disk for the fiducial model. While these parameterizations aim to reproduce some of the general characteristics of the thermal variations derived for disk models that incorporate annular gaps or radial drift, simplifications are also made (e.g., the temperature step function in the inner disk) in order to allow for a more direct comparison with the fiducial model. The model parameters are listed in Table \ref{tab:models}. 

As shown in Figures \ref{fig:modelimages} and \ref{fig:insets}, the emission morphology within a radius of 150 AU is quite similar to the observations. As expected for optically thick CO, the relative variations in the thermal profile necessary to create emission substructure are much smaller than the variations that would be required in the column density profile. Whereas the key observed emission features were reproduced in the fiducial model through a factor of 40 decrease in the CO column density at 15 AU and then a secondary bump by a factor of a few in the outer disk, the ``perturbed temperature" model only boosts temperatures by 60\% ($\delta T = 1.6$) in the inner disk and by 15\% ($B = 0.15$) at the peak of the secondary ring relative to the fiducial model. The width of $g_2(r)$, setting the secondary bump in the temperature profile, is somewhat wider than that of $f_2(r)$, which sets the secondary bump in the CO column density profile in the fiducial model. Since the temperature profile declines more steeply with radius than the CO column density profile, the temperature has to be increased over a wider region to reproduce the same emission bump that is generated by an increase in the CO column density.

While a smooth CO surface density profile is used to isolate the effects of temperature in creating emission substructure, this model is not physically self-consistent. Although the ``perturbed temperature" model suggests that the depletion and enhancement factors in the CO surface density profile may be more modest than those used for the fiducial model, the mechanisms that have been proposed for creating localized thermal variations should still yield variations in the CO column density. For example, \citet{2012ApJ...749..153J} find that a disk gap with the surface density reduced by a factor of 6 can yield thermal variations on the order of $25\%$ compared to a smooth disk. For thermal variations mediated by radial drift, heating of the outer disk would promote CO ice desorption, thereby creating a secondary CO column density bump. Hence, while the CO emission substructure may be largely a temperature effect from the radiative transfer point of view, they would still ultimately signify the presence of surface density substructure.

\section{Discussion}\label{sec:discussion}

\subsection{Possible Origins of the CO Emission Features}
Our fiducial model indicates that the emission morphology of $^{12}$CO in the TW Hya disk can be reasonably reproduced with a steep decrease in the $^{12}$CO column density at a radius of $\approx$ 15 AU, followed by a secondary peak at a radius of $\approx$ 70 AU. To evaluate what scenarios are likely to have created these column density variations, we consider the $^{12}$CO results in the context of other observations of the TW Hya disk as well as physical and chemical modeling results from the literature.  

\subsubsection{Midplane CO freezeout}
Estimates for the TW Hya disk's midplane CO snowline location range from 11 to 33 AU \citep{2013Sci...341..630Q, 2016ApJ...819L...7N,2016ApJ...823...91S, 2017ApJ...840...93P, 2017AA...599A.101V,2017NatAs...1E.130Z}, so it is natural to consider whether the apparent steep drop in $^{12}$CO column density at a radius of 15 AU is related. Indeed, several of the aforementioned CO snowline estimates are based on observations of a dip in the C$^{18}$O emission profile at a radius (22 AU) close to where we infer a $^{12}$CO column density drop. 

As noted in our analysis of the $^{12}$CO brightness temperatures in \ref{sec:nofreezeout}, the $^{12}$CO emission appears to originate well above the CO snow surface, which is in line with typical assumptions about disks. This implies that the inferred column density drop at $r=15$ AU and the bump at $r=70$ AU do not directly trace the onset of CO freezeout in the midplane. Because similar emission features are observed in $^{13}$CO and C$^{18}$O $J=3-2$ \citep{2016ApJ...819L...7N,2016ApJ...823...91S, 2017NatAs...1E.130Z}, which emit from different heights in the disk due to their lower optical depths, we argue that the line observations together likely trace CO depletion or gas density reductions occurring throughout the vertical extent of the warm molecular layer, not just near the midplane. Consequently, while the aggregate evidence indicates that the midplane CO snowline does lie somewhere between 11 and 33 AU in the TW Hya disk, we advise general caution in using CO isotopologue observations to infer the CO snowline location.

\subsubsection{Gas surface density substructures\label{sec:depressions}}
Optically thick $^{12}$CO emission does not directly constrain the molecular gas distribution of the TW Hya disk, but the distribution of sub-micron-sized dust grains has been used as a proxy because  small grains are usually well-coupled to gas \citep{2013ApJ...771...45D, 2015ApJ...802L..17A, 2015ApJ...815L..26R, 2017ApJ...837..132V}.  Most recently, \citet{2017ApJ...837..132V} presented SPHERE scattered light observations showing wide radial depressions at $\approx23$ and $\approx94$ AU, which they interpreted as tracing gas surface density variations. These depressions overlap with where we infer steep $^{12}$CO column density decreases in our fiducial model, which may indicate that the $^{12}$CO emission is also following gas surface density variations.

Nonetheless, there are at least two apparent discrepancies between the inferred gas surface density profile from \citet{2017ApJ...837..132V} and the inferences we have made regarding the CO distribution in the TW Hya disk. First, their inferred gas surface density profile attains a local maximum just outside $r=100$ AU, whereas none of the CO isotopologue observations show an obvious rise in intensity outside this radius. Complementary chemical modeling of double CS emission rings by \citet{2017ApJ...835..228T} indicates that their observations are best reproduced by a gas surface density depression similar to the one inferred by \citet{2017ApJ...837..132V} at a radius of 94 AU. One possible resolution to this discrepancy is that the small dust grains become less well-coupled to the gas in the more tenuous upper layers of the cold outer disk. Alternatively, CO traces the gas surface density decrease into the 94 AU gap, but does not rise again with the gas surface density due to more substantial CO freezeout in the outer disk.

Second, the relative amplitudes of the gas surface density variations inferred by \citet{2017ApJ...837..132V} range roughly from 45 to 80\%. The \citet{2017NatAs...1E.130Z} model, which motivated our fiducial model, derived a C$^{18}$O  column density drop by an order of magnitude in the inner disk and then an increase by a factor of a few to create the secondary ring. While our ``perturbed temperature" model provides an example of how the $^{12}$CO column density variations needed to reproduce the observations are sensitive to the prescribed temperature structure, column density estimates from the rarer and therefore less optically thick isotopologues should be more robust. 

On the other hand, temperature, dust surface density, and dust opacities are degenerate with one another, so more stringent constraints on the temperature structure of the TW Hya disk would be essential to determine whether more extreme gas surface density variations are also compatible with the scattered light data. In the meantime, we do not rule out the possibility that substructure in the underlying gas disk at least partially contributes to the observed CO emission morphology.

\subsubsection{CO Depletion in Warm Gas\label{sec:depletion}}
Another possibility to consider is that the features are due to spatial variations of $X_\text{CO}$ in warm gas: $X_\text{CO}$ drops by one to two orders of magnitude near 15 AU and then rises by a factor of a few in the outer disk before decreasing again, creating the secondary CO column density bump peaking at $\approx 70$ AU. While several studies have suggested that warm gas in the TW Hya disk is CO-depleted by one to two orders of magnitude relative to the ISM \citep{2013ApJ...776L..38F, 2016AA...592A..83K, 2016ApJ...819L...7N, 2016ApJ...823...91S, 2017ApJ...840...93P}, the new high angular resolution ALMA observations provide additional insight into how the CO distribution varies radially.

Destruction of CO by He$^+$ is an oft-proposed mechanism for CO depletion in disks \citep[e.g.][]{2013ApJ...776L..38F, 2016ApJ...819L...7N, 2016ApJ...822...53Y}. Stellar X-rays ionize helium, which then reacts with CO \citep{1973ApJ...185..505H, 2014FaDi..168...61B}:
\begin{equation}
\text{He}^+ + \text{CO} \rightarrow \text{C}^+ +\text{O} +\text{He}. 
\end{equation}
Much of the C$^+$ is thought to be incorporated back into CO, but alternative pathways incorporating  C$^+$ into CO$_2$, hydrocarbons, and complex oxygen-bearing molecules can still  lead to significant CO depletion over timescales of several million years \citep[e.g.][]{1997ApJ...486L..51A, 2014FaDi..168...61B, 2016ApJ...822...53Y}. 

Alternatively, CO depletion in the warm molecular layer could be directly tied to CO freezeout deeper in the disk. \citet{2017ApJ...835..162X} presented ``turbulent-diffusion mediated CO depletion" models demonstrating that if small dust grains are reasonably settled in a disk with weak turbulence, then the fractional abundance of gas-phase CO in the warm molecular layer can eventually be reduced by an order of magnitude as CO diffuses into the cold midplane and then freezes out. However, turbulence limits for the inner disk of TW Hya would have to be obtained to assess the feasibility of this mechanism.

Conversely, rather than interpreting the disk outside 15 AU as being CO-depleted in the warm molecular layer, the inner disk may be enhanced in CO if ice-coated particles drift inward and the ice subsequently sublimates \citep[e.g.][]{2004ApJ...614..490C, 2017AA...600A.140S}. The models investigating this effect have so far been one-dimensional. Further study on the heights at which this abundance enhancement occurs would help to determine the impact on CO emission, since enhancement limited to the midplane would not make much difference to optically thick emission. 

While gas density substructures and CO depletion have heretofore been discussed as separate possibilities, it is worth considering the extent to which they may be coupled. Reduced gas and dust surface densities lead to reduced UV opacities, which may allow UV radiation to dissociate CO more easily within the disk gaps \citep[e.g.][]{2009AA...503..323V}. In addition, bumps and dips in the gas surface density profile can regulate dust transport and growth \citep[e.g.][]{1972fpp..conf..211W, 2012AA...538A.114P}, which in turn may affect the efficiency of the grain surface reactions that serve as carbon sinks. Further chemical modeling would be useful to examine the extent to which CO depletion can be mediated by gas and dust gaps.

\subsection{Possible Origins of the Radial Spectral Index Variations}
We discuss below potential explanations for the appearance of the TW Hya continuum spectral index map, which shows a striking pattern of low values at the bright emission rings and high values within the emission gaps. 

\subsubsection{Radial Variation in Grain Sizes}
Radial variations in $\alpha$ are often attributed to spatially varying grain size distributions in protoplanetary disks \citep[e.g.][]{2012ApJ...760L..17P, 2014AA...564A..93M, 2016ApJ...829L..35T}. Previous multi-frequency measurements of TW Hya have indicated the presence of centimeter-sized grains inward of $r \sim 15$ AU and the exclusion of such large grains from the 25 AU continuum gap \citep{2005ApJ...626L.109W, 2014AA...564A..93M, 2016ApJ...829L..35T}. With the high angular resolution 290 GHz spectral index map, we can also improve constraints on the grain size distribution in the narrow gaps and bright rings outside 25 AU.

Following \citet{2016ApJ...829L..35T}, we estimate the intensity of the submillimeter dust emission as 
\begin{equation}\label{eq:intensity}
I_\nu (r) = B_\nu(T_d(r))(1-\exp[-\tau_\nu(r)]),
\end{equation}
where $B_\nu$ is the Planck function, $T_\text{d}(r)$ is the dust temperature (assuming that the millimeter dust disk is essentially isothermal vertically), and $\tau_\nu(r)=\tau_\text{290 GHz}(r)(\nu/\text{290 GHz})^\beta$ is the dust optical depth. The spectral index is then
\begin{equation}\label{eq:alpha}
\alpha(r) = 3-x \frac{\exp{x}}{\exp{x}-1}+\beta(r) \frac{\tau_\nu(r)}{\exp{\tau_\nu(r)}-1},
\end{equation}
where $x = \frac{h\nu}{k_B T_\text{d}(r)}$.  

The three unknowns are $\beta$, $T_d$, and $\tau_\nu$. While there are not sufficient constraints to solve for all three unknowns, useful limits can be placed on $\beta$. Figure \ref{fig:alpha} shows dust temperatures as a function of brightness temperatures for different fixed values of $\tau$, and then $\alpha$ as a function of $\tau$ and $\beta$ for a fixed brightness temperature of 11 K. The value of 11 K is chosen based on the brightness temperatures of the 290 GHz continuum emission rings at 35 and 45 AU. Because $\beta$ decreases as the maximum grain size increases, the upper limit on $\beta$ should be the ISM value of $\approx1.7$ \citep{2006ApJ...636.1114D}, when no grain growth has occurred. Since $\alpha$ decreases when $\tau$ increases and when $\beta$ decreases, placing a lower limit on $\tau$ also places a lower limit on $\beta$ when $\alpha$ is known.

The dust temperature is likely no more than 30 K in the gaps and rings beyond a radius of 25 AU in the TW Hya, given that CO freezeout is expected to commence in the midplane somewhere between 11 and 33 AU \citep[e.g.][]{2013Sci...341..630Q, 2016ApJ...819L...7N, 2017NatAs...1E.130Z}. The brightness temperatures at radii from 25 to 50 AU range from 10 to 12 K. Using Equation \ref{eq:intensity}, this suggests that $\tau\gtrapprox 0.25$ at 290 GHz in the gaps and rings outside 25 AU.

The peak $\alpha$ values at the 25, 41, and 47 AU continuum emission gaps are between 2.6 and 2.7. A lower bound of $\tau=0.25$ indicates that $\beta\gtrapprox 1$ in order to reproduce these $\alpha$ values, which would allow for grain growth up to a few mm \citep{2006ApJ...636.1114D}. In contrast, the low $\alpha$ values measured at the bright emission rings at 35 and 45 AU are compatible with $\beta\gtrapprox0.4$, which allows for centimer-size grains \citep[e.g.][]{2012ApJ...760L..17P}. 

\citet{2016ApJ...829L..35T} interpreted the high spectral index value in the 25 AU continuum gap as a signature of an embedded planet creating a pressure gradient that allows small grains to enter the gap but excludes large grains. This mechanism, known as dust filtration, is modeled in detail by \citet{2006MNRAS.373.1619R}, \citet{2006AA...453.1129P}, \citet{2012ApJ...755....6Z},  and \citet{2012AA...545A..81P}, among others. Dust filtration may also explain the gaps at 41 and 47 AU, given their similar spectral index signatures. 

\subsubsection{Optical Depth Variations}
While spatially varying grain size distributions can generate the observed spectral index profile, an alternative scenario worth considering is that the radial spectral index variations are largely tracing optical depth variations instead. Previous analyses of millimeter continuum emission concluded that $\tau\sim1$ at $r<15$ AU in the TW Hya disk, in part based on the low spectral index values measured in the inner disk \citep{2016ApJ...820L..40A, 2016ApJ...829L..35T, 2017NatAs...1E.130Z}. However, the high-resolution 290 GHz spectral index radial profile also reveals low spectral index values of $\alpha\approx 2$ at the locations of the bright emission rings at $r\approx35$ and $r\approx45$ AU, raising the question of whether these are also marginally optically thick. We note that given the systematic uncertainties discussed in Section \ref{sec:results}, it is possible that the true $\alpha$ profile is shifted upward by up to $\Delta\alpha = 0.4$, which would imply that nearly the whole millimeter dust disk is optically thin. For the sake of simply formulating a plausibility argument for the optically thick scenario, though, the remainder of the calculations in this section will take the spectral index profile in Figure \ref{fig:spix} at face value.

\begin{figure*}[htp]
\centering
\includegraphics[scale = 0.8]{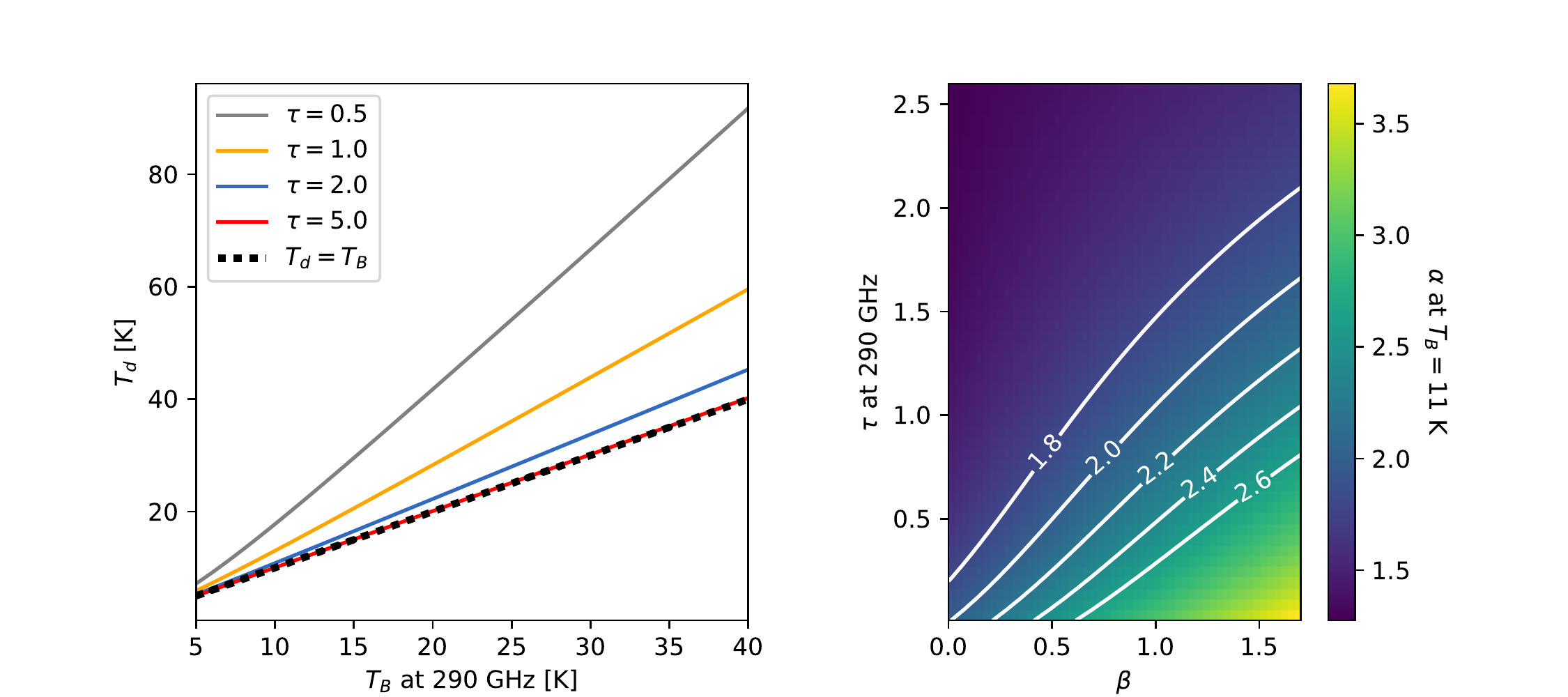}

\caption{\textit{Left}: Plot of the relationship between $T_B$ and $T_d$, computed at $\nu=290$ GHz and for several values of $\tau$. \textit{Right}: Plot of $\alpha$ at $T_B=11$ K as a function of $\tau$ and $\beta$.}
\label{fig:alpha}
\end{figure*}

Referring again to Figure \ref{fig:alpha}, two points are worth emphasizing. First, low brightness temperatures are sometimes taken as prima facie evidence that the dust is optically thin, since the brightness temperature and dust temperature should be equal at large optical depths. The first panel of Figure \ref{fig:alpha} illustrates, though, that even at an optical depth as high as $\tau = 1$, the brightness temperature is $\sim30\%$ lower than the corresponding dust temperature. Thus, while \citet{2017NatAs...1E.130Z} point out that adopting the brightness temperature as the midplane dust temperature would yield an unrealistically cold disk, a value of $\tau = 1$ corresponds to midplane temperatures that are only a few degrees lower than previous TW Hya models \citep[e.g.][]{2013Sci...341..630Q, 2017NatAs...1E.130Z}. Second, $\alpha = 2$ is sometimes quoted as the lower bound set by optically thick emission, but this is specific to the Rayleigh-Jeans limit. The second panel of Figure \ref{fig:alpha} illustrates that at millimeter wavelengths, emission can still be optically thin when $\alpha = 2$, provided that $\beta$ is small (and therefore grain sizes are large). For a dust temperature of 11 K, which can occur in the outer regions of protoplanetary disks, $\alpha$ can be as low as 1.2 in the optically thick limit at 290 GHz. 

Figure \ref{fig:spix} shows that the azimuthally averaged spectral index values range from roughly 2.1 to 2.3 inside the bright emission rings at 35 and 45 AU. These $\alpha$ values are possible if $\tau\lessapprox1.5$ (see Figure \ref{fig:alpha}), still allowing the rings to be marginally optically thick.  Even higher local values of $\tau$ may be compatible with the low brightness temperatures if the bright emission rings at 35 and 45 AU are actually a series of unresolved rings that result in a low beam filling factor \citep[e.g.][]{2012AA...540A...6R}. \citet{2016ApJ...820L..40A} tentatively identified additional emission gaps between 30 and 35 AU, noting that higher angular resolution and sensitivity would be needed to confirm. 

High dust optical depths imply high surface densities, so the gravitational stability of the potential unresolved narrow rings should be considered. The stability of the rings can be crudely estimated with Toomre's $Q$ parameter \citep{1964ApJ...139.1217T}, where $Q = c_s \Omega /\pi G \Sigma$. (This estimate does require the assumption of a geometrically thin disk, which may not necessarily be true for TW Hya.) Assuming an ISM gas-to-dust ratio of 100, $\Sigma\approx 100 \Sigma_\text{dust} \approx 100 \tau_\text{dust}/\kappa_\text{dust}$. We adopt a dust opacity at 290 GHz of $\kappa = 3.3$ cm$^2$ g$^{-1}$, based on the Mie scattering calculations reported in \citet{2009ApJ...700.1502A} for spherical grains with a minimum radius of $a=0.005$ $\mu$m, a maximum radius of 1 mm, a power-law distribution of $n(a)\propto a^{-3.5}$, silicate and graphite abundances derived in \citet{1977ApJ...217..425M}, and dielectric functions from \citet{1984ApJ...285...89D} and \citet{2001ApJ...548..296W}. For a stellar mass of 0.88 $M_\odot$ and an isothermal sound speed calculated at a temperature of 15 K, $Q>1$ for $\tau_\text{dust}<3.2$ at 35 AU and for $\tau_\text{dust}<2.2$ at 45 AU, suggesting that the rings can be optically thick and remain marginally gravitationally stable. Our calculation for $Q$ scales with the ratio $\kappa_\text{dust}/\tau_\text{dust}$. Based on a distribution-of-hollow-spheres (DHS) calculation for carbonaceous silicates, \citet{2016AA...586A.103W} advocate for the use of higher dust opacity values compared to the typically used Mie scattering values. Such higher opacity values would allow higher dust optical depths while maintaining gravitational stability. 
While optical depth effects may largely be sufficient to explain the spectral index at the gaps and rings between 25 and 45 AU, the final rise in the spectral index near the 47 AU emission gap is probably at least partially due to radial changes in $\beta$, since the peak in the spectral index profile is shifted slightly outward from the 47 AU gap. The continuum intensity depends on temperature and optical depth, so if the radial temperature gradient is not too steep, the radial locations of the minima in the emission profiles would be close to the locations of the minima in the optical depth profiles. On the other hand, $\alpha$ has an additional dependence on $\beta$. The 47 AU emission gap is not far from the edge of the millimeter dust disk, so the nearby rise in the spectral index may be influenced both by the deficit of material within the gap itself and by the bulk loss of large dust grains in the outer disk.   

\subsubsection{Comparison to Other Sources}
Our calculations raise the possibility that all of the bright rings in the TW Hya millimeter continuum are marginally optically thick. This possibility has been suggested for other disks as well\textemdash to explain the correlation between the continuum luminosity and emitting area for a large disk sample, \citet{2017ApJ...845...44T} proposed that these disks consist of narrow, optically thick rings. Dust masses are often estimated using the \citet{1990AJ.....99..924B} formula assuming optically thin millimeter disk emission. If most disks are made up of optically thick rings, then dust masses have been systematically underestimated. In this case, CO or gas depletion may be even more severe than currently estimated, although large line optical depths may also contribute to underestimates of the amounts of CO or gas \citep[e.g.][]{2016ApJ...828...46A, 2016ApJ...827..142B, 2017ApJ...844...99L}. High-resolution, multi-band ALMA surveys of protoplanetary disks can determine whether other disks have continuum emission and spectral index profiles similar to that of TW Hya.  

Only one other protoplanetary disk, that of HL Tau, has had its millimeter spectral index mapped at an angular resolution comparable to the TW Hya observations  \citep{2015ApJ...808L...3A}. The features are strikingly similar to those of TW Hya\textemdash the bright dust emission rings correspond to low spectral index values of $\alpha\approx2$ and the emission gaps correspond to high values of $\alpha\approx 3$. HL Tau and TW Hya may therefore reflect a common pattern in disks. If this is the case, the abrupt changes in the radial spectral index profiles may correspond to characteristic width scales for disk gaps. Measuring gap widths from intensity profiles is associated with significant ambiguity because an underlying ``unperturbed" profile has to be assumed \citep[e.g.][]{2016ApJ...818..158A, 2016PASJ...68...43K}. Hydrodynamic models of planet-forming disks often predict gap widths as a function of planet mass \citep[e.g.][]{2010AA...518A..16F, 2013ApJ...769...41D, 2013ApJ...768..143Z,  2015ApJ...809...93D, 2016ApJ...818...76J}, so linking gap widths in physical models to the widths of rings in resolved spectral index profiles may be useful for placing upper limits on the masses of potential embedded planets. However, it is currently ambiguous which, if any, features in the TW Hya disk are associated with embedded planets. While recent hydrodynamic simulations suggest that a single low-mass planet can create multiple gaps reminiscent of those observed in the millimeter continuum emission of TW Hya \citep{2017ApJ...835..146D, 2017ApJ...850..201B}, \citet{2017ApJ...837..132V} note that their derived gas surface density profile differs from predictions of gaps opened by planets in \citet{2015ApJ...807L..11D}.

\subsection{Improving constraints on the structure of the TW Hya disk}
While relatively simple parameterizations of the molecular abundance and temperature structures can reasonably reproduce the radial features observed in the single  $^{12}$CO transition available at high resolution, better constraints need to be obtained on the vertical structure. Observing multiple transitions of all the isotopologues at comparably high angular resolution can reduce uncertainties with respect to the vertical distribution of CO. For example, whereas the inner edge of our $^{12}$CO surface density gap lies at a radius of 15 AU, \citet{2017NatAs...1E.130Z} constrained the inner edge of the gap to lie at a radius of 22 AU for C$^{18}$O. This modest difference may be due to the coarser angular resolution of the C$^{18}$O observations, but alternatively it could result from the onset radius of CO or gas depletion varying slightly with height, or from isotope-selective effects. Higher-resolution observations of C$^{18}$O would clarify the nature of this apparent difference. Several works have also suggested that the inner few AU of the TW Hya disk may either be warped or feature deviations from Keplerian rotation \citep[e.g.][]{2005ApJ...622.1171R, 2012ApJ...757..129R, 2017ApJ...835..205D}. These effects alone cannot create the axisymmetric substructures observed in the main CO isotopologues, but they would be worth considering in a more detailed analysis of the disk vertical structure. 

With the observations currently available, it is not straightforward to distinguish between gas disk substructure and CO depletion as the main explanation for the CO emission morphology. Although the gas surface density profile derived by \citet{2017ApJ...837..132V} appears to disfavor the possibility that CO is closely tracing the gas distribution, they assumed a constant gas-to-small-dust ratio. TW Hya is one of the few sources for which a disk gas mass has been measured without having to assume either an $X_\text{CO}$ value or gas-to-dust ratio, thanks to \textit{Herschel} observations of HD \citep{2013Natur.493..644B}. Although \citet{2015ApJ...799..204C} find that the bulk gas-to-dust ratio of the TW Hya disk is consistent with that of the ISM, the spatially unresolved HD observations do not exclude localized gas depletions. 

The question of how to reliably distinguish CO and gas depletion in disks has been raised in a number of works aiming to estimate the gas masses of protoplanetary disks from millimeter/submillimeter surveys of CO and dust  \citep[e.g.][]{ 2016ApJ...827..142B,2017ApJ...844...99L, 2017AA...599A.113M}. As the nearest disk, TW Hya may be the most important test case for breaking the degeneracy. \citet{2017AA...599A.113M} suggest measuring hydrocarbon abundances to check for evidence of CO depletion, since hydrocarbons are believed to be a carbon sink when CO is destroyed by He$^+$. C$_2$H and $c$-C$_3$H$_2$ have been observed in the TW Hya disk at $\approx0\farcs4$ resolution by \citet{2016ApJ...831..101B}; re-observing at higher resolution to compare to CO emission, accompanied by chemical modeling, may yield additional insight into whether and how $X_\text{CO}$ varies throughout the disk. 

(Magneto)hydrodynamic simulations tailored to TW Hya would also shed some light on whether the CO emission morphology is tracing gas substructure rather than chemical depletion. While a direct correspondence between the CO and millimeter dust structures is not obvious in the data presented in this work, disk models indicate that embedded planets and dead zones can create gaps and rings in gas and millimeter dust that are offset from one another and differ in depth and width \citep[e.g.][]{2016AA...596A..81P, 2016AA...590A..17R, 2017ApJ...843..127D,2017arXiv171004418F}. A model that simultaneously matches the TW Hya millimeter observations and yields a gas surface density profile similar to that of the inferred CO surface density profile would be compelling evidence for gas substructure.

\section{Summary}\label{sec:summary}
We presented new ALMA observations of $^{12}$CO $J=3-2$ in the TW Hya disk at a spatial resolution of 8 AU, representing one of the highest resolution images so far of molecular line emission in a protoplanetary disk. We also reprocessed archival 1.3 mm and 870 $\mu$m continuum data to produce a spectral index map at a spatial resolution of 2 AU. Our results and conclusions are as follows: 
\begin{enumerate}
\item The $^{12}$CO images exhibit radial emission breaks coinciding with gaps and rings previously observed in $^{13}$CO and C$^{18}$O emission. We employ LTE radiative transfer modeling to demonstrate that the $^{12}$CO emission morphology can be reasonably reproduced by a sharp drop in the CO column density at $r=15$ AU and a secondary peak at 70 AU. 
\item Analysis of the $^{12}$CO brightness temperatures and radiative transfer modeling suggest that the inferred CO column density variations are likely not associated with the onset of freezeout in the midplane. We propose instead that the variations in the $^{12}$CO column density arise either from spatial variations in $X_\text{CO}$ or from gas density reductions in the warm molecular layer of the disk. Based on similar features observed in the much lower abundance isotopologues, we further argue that these variations are likely present throughout most of the vertical extent of the warm layer. Distinguishing between CO depletion and gas disk substructures would be facilitated by obtaining more stringent constraints on the temperature structure of the disk and observing complementary molecular tracers of carbon depletion. 
\item The 290 GHz spectral index map shows a striking contrast between spectral index values of $\sim2$ at the bright continuum emission rings and $\sim2.7$ at the emission gaps. The high spectral index values within the emission gaps suggest that the maximum grain size is limited to a few millimeters. The low spectral index values at the continuum emission rings may be a signature of grain growth to centimer sizes, but a plausible alternative explanation is that the rings are all marginally optically thick. The latter possibility is worth investigating for a larger sample of disks to determine whether disk dust masses are being systematically underestimated. 

\end{enumerate}
\acknowledgments
We thank the NAASC staff for their advice on data reduction; Roy van Boekel, Hannah Jang-Condell, and Ke Zhang for their discussions of TW Hya; and the referee for useful comments. This paper makes use of ALMA data \dataset[ADS/JAO.ALMA\#2012.1.00422.S]{https://almascience.nrao.edu/aq/?project\_code=2012.1.00422.S}, 
\dataset[ADS/JAO.ALMA\#2013.1.00114.S]{https://almascience.nrao.edu/aq/?project\_code=2013.1.00114.S}, 
\dataset[ADS/JAO.ALMA\#2013.1.00196.S]{https://almascience.nrao.edu/aq/?project\_code=2013.1.00196.S}, \dataset[ADS/JAO.ALMA\#2013.1.00198.S]{https://almascience.nrao.edu/aq/?project\_code=2013.1.00198.S}, \dataset[ADS/JAO.ALMA\#2013.1.00387.S]{https://almascience.nrao.edu/aq/?project\_code=2013.00387.S}, \dataset[ADS/JAO.ALMA\#2013.1.01397.S]{https://almascience.nrao.edu/aq/?project\_code=2013.1.01397.S}, \dataset[ADS/JAO.ALMA\#2015.1.00686.S]{https://almascience.nrao.edu/aq/?project\_code=2015.1.00686.S}, \dataset[ADS/JAO.ALMA\#2015.A.00005.S]{https://almascience.nrao.edu/aq/?project\_code=2015.A.00005S}, and \dataset[ADS/JAO.ALMA\#2016.1.00629.S]{https://almascience.nrao.edu/aq/?project\_code=2016.1.00629.S}. ALMA is a partnership of ESO (representing its member states), NSF (USA) and NINS (Japan), together with NRC (Canada) and NSC and ASIAA (Taiwan), in cooperation with the Republic of Chile. The Joint ALMA Observatory is operated by ESO, AUI/NRAO and NAOJ. The National Radio Astronomy Observatory is a facility of the National Science Foundation operated under cooperative agreement by Associated Universities, Inc. J.H. acknowledges support from the National Science Foundation Graduate Research Fellowship under Grant No. DGE-1144152 and from NRAO Student Observing Support. LIC acknowledges the support of NASA through Hubble
Fellowship grant HST-HF2-51356.001-A awarded by the Space Telescope Science Institute, which is operated by the Association of Universities for Research in Astronomy, Inc., for NASA, under contract NAS 5-26555. KI\"O acknowledges funding through a Packard Fellowship for Science and Engineering from the David and Lucile Packard Foundation. T.B. acknowledges funding from the European Research Council (ERC) under the European Union's Horizon 2020 research and innovation programme under grant agreement No 714769.

\software{\texttt{CASA} \citep{2007ASPC..376..127M}, \texttt{AstroPy} \citep{2013AA...558A..33A}, \texttt{analysisUtils} (\url{https://casaguides.nrao.edu/index.php/Analysis_Utilities}), \texttt{RADMC-3D} \citep{2012ascl.soft02015D}, \texttt{radmc3dPy} (\url{http://www.ast.cam.ac.uk/~juhasz/radmc3dPyDoc/index.html}), \texttt{vis\_sample} \citep{2015ApJ...806..154C, loomis}}
\facility{ALMA}

\bibliographystyle{aasjournal}
\bibliography{twhya}
\appendix

\section{Additional Calibrator Details}

\begin{deluxetable*}{cccc|cccc}
\tablecaption{Calibrator Sources and Calibration Models\label{tab:fluxcal}}
\tablehead{
\colhead{Program}&\colhead{Date(s)} &\colhead{Bandpass}&\colhead{Phase}&\colhead{Flux} &\colhead{Reference Frequency}&\colhead{Flux Density}&\colhead{Spectral Index}\\
&&calibrator&calibrator&calibrator&\colhead{(GHz)}&\colhead{(Jy)}}
\startdata
2012.1.00422.S & 2015 May 14 &J1256$-$057&J1037$-$2934&Titan\tablenotemark{a}&-&-&-\\
2013.1.00114.S & 2014 July 19&J1037$-$2934&J1037$-$2934&Pallas\tablenotemark{a}&-&-&-\\
2013.1.00196.S& 2014 Dec. 24 &J1256$-$0547&J1037$-$2934&Callisto\tablenotemark{a}&-&-&-\\
& 2015 April 5 &J1256$-$0547&J1037$-$2934&J1037$-$295&330.624&0.736&-0.53\\
2013.1.00198.S&2014 Dec. 31 &J1058+0133&J1037$-$2934&Ganymede\tablenotemark{a}&-&-&-\\
&2015 June 15 &J1058+0133&J1037$-$2934&Callisto\tablenotemark{a}&-&-&-\\
2013.1.00387.S & 2015 May 13 &J1037$-$2934&J1037$-$2934&Ganymede\tablenotemark{a}&-&-&-\\
2013.1.01397.S&2015 May 19-20&J1058+0133&J1037$-$2934&J1037$-$295 &335.726& 0.587 &-0.534\\
2015.1.00686.S &2015 Nov. 23&J1058+1033&J1103$-$3251&J1037$-$2934 &350.638&0.604&-0.492\\
&2015 Nov. 30 &J1058+0133&J1103$-$3251&J1107$-$4449&350.640&0.542&-0.711\\
&2015 Dec. 1&J1058+0133&J1103$-$3251&J1037$-$2934&350.640&0.627&-0.468\\
2015.A.00005.S&2015 Dec. 1&J1107$-$4449&J1103$-$3251&J1037$-$2934 & 233.0& 0.759 & -0.468\\
2016.1.00629.S&2016 Dec. 30&J1058+0133&J1037$-$2934 & J1037$-$2934 \tablenotemark{b} & 345.810& 0.728 & -0.357\\
&2016 Dec. 30& J1058+0133&J1037$-$2934&J1058+0133\tablenotemark{b} & 345.809 & 3.568&-0.395\\
&2017 July 4 & J1037$-$2934 & J1037$-$2934 & J1037$-$2934 &345.755 & 0.788 &-0.592 \\
&2017 July 9 & J1058+0133 & J1037$-$2934 & J1037$-$2934&345.756 & 0.788 &-0.592\\
& 2017 July 14 & J1058+0133 & J1037$-$2934 & J1037$-$2934&345.756 & 0.646 & -0.647 \\
& 2017 July 20 & J1058+0133 & J1037$-$2934 & J1037$-$2934& 345.758 & 0.646 & -0.647\\
& 2017 July 21 & J1058+0133 & J1037$-$2934 & J1256$-$0547 & 345.757 & 7.4223 & -0.495\\ 
\enddata
\tablenotetext{a}{Used the Butler-JPL-Horizons 2012 models}
\tablenotetext{b}{Flux density values adjusted from original delivered scripts to correspond with updated calibrator catalog values. All other values are retained from the ALMA archival scripts.}
\end{deluxetable*}
Table \ref{tab:fluxcal} lists the bandpass, phase, and flux calibrators used for all the data presented in this work, as well as the reference frequency, flux density, and spectral index used to model the spectra of the quasars that served as flux calibrators. The ALMA calibrator catalog flux density values that served as the basis for the flux calibration on 2016 December 30 for program 2016.1.00629.S were revised downward by 8$\%$ after delivery of the original data (T. Hunter, private communication), so we rescaled the fluxes of the datasets accordingly. In all other cases, we retained the original flux calibration specified in the ALMA archive. 

\section{Band 6 and 7 Continuum Images and Profiles} \label{sec:continuum}

\begin{figure*}[htp]
\centering
\includegraphics[scale = 0.6]{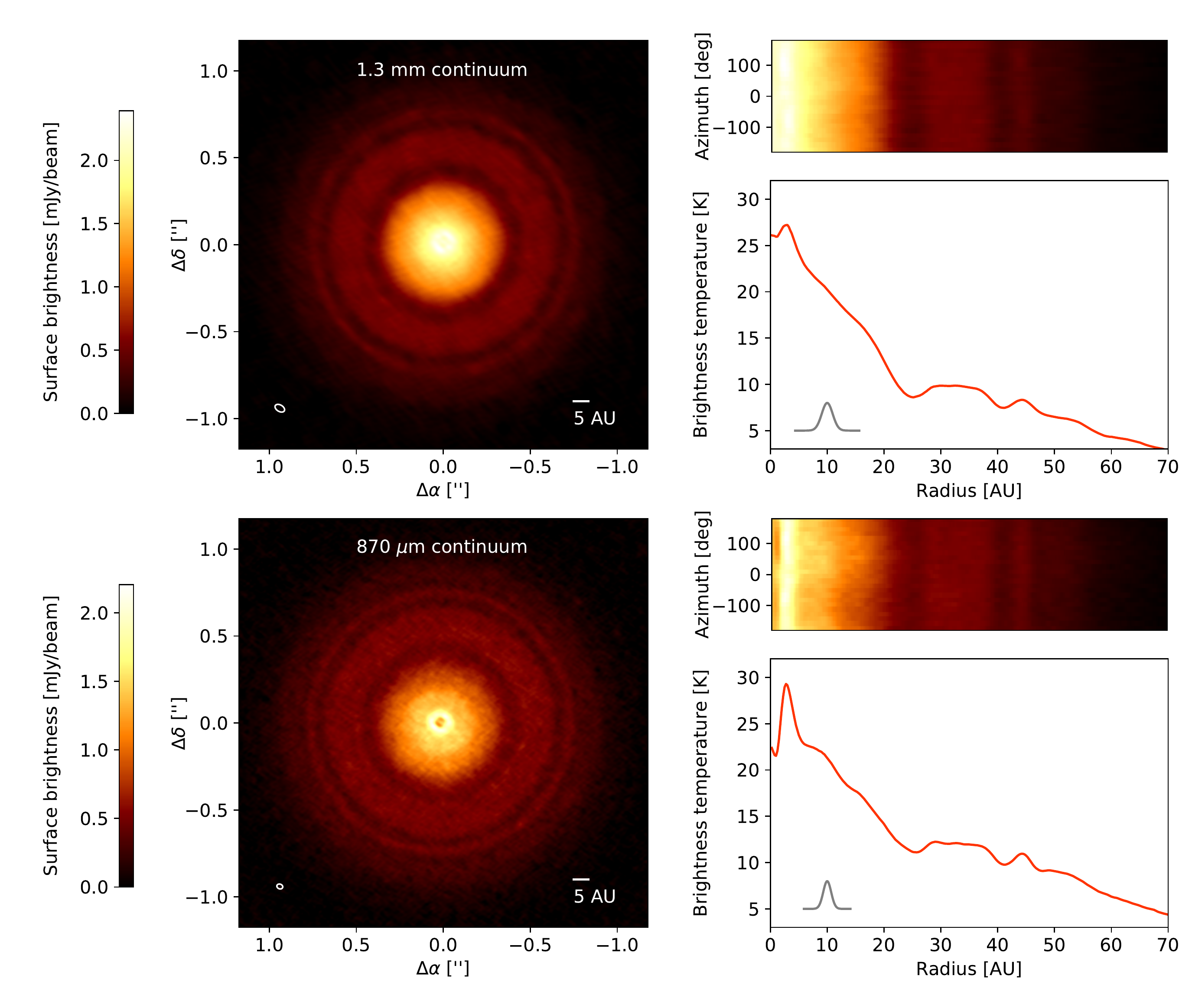}

\caption{\textit{Top left}: Image of 1.3 mm continuum emission. The synthesized beam, shown in the lower left corner, has FWHM dimensions of $61\times38$ mas (3.6 $\times$ 2.2 AU) and a position angle of 59$\fdg$4. The rms is 18 $\mu$Jy beam$^{-1}$. \textit{Top right}: The 1.3 mm continuum emission deprojected and replotted as a function of radius and azimuth, with the deprojected, azimuthally averaged radial brightness temperature profile shown underneath. The Gaussian profile shows the FWHM of the minor axis of the synthesized beam. \textit{Bottom left}:  Image of 870 $\mu$m continuum emission. The synthesized beam has FWHM dimensions of $35\times28$ mas (2.1 $\times$ 1.7 AU) and a position angle of 69$\fdg$9. The rms is 31 $\mu$Jy beam$^{-1}$. \textit{Bottom right}: The 870 $\mu$m continuum emission plotted in polar coordinates, with the corresponding averaged radial brightness temperature profile underneath.}
\label{fig:individualcont}
\end{figure*}

The 1.3 mm and 870 $\mu$m continuum images and radial brightness temperature profiles are shown in Figure \ref{fig:individualcont}. The inclusion of additional archival data provides a modest improvement in image fidelity over the images from \citet{2016ApJ...820L..40A} and \citet{2016ApJ...829L..35T}. The Planck equation is used to compute brightness temperature because the Rayleigh-Jeans approximation is typically poor for submillimeter observations of protoplanetary disks. 
\section{$^{13}$CO $J=3-2$ Channel Maps}\label{sec:13CO}
For comparison with the $^{12}$CO data, we image $^{13}$CO $J=3-2$ in the TW Hya disk by combining archival observations from programs 2012.1.00422.S, 2013.1.00196.S, and 2013.1.01397.S. Integrated intensity maps of $^{13}$CO from 2012.1.00422.S and 2013.1.01397.S were presented separately in \citet{2016ApJ...823...91S} and \citet{2016ApJ...819L...7N}, respectively. The raw data were downloaded from the ALMA archive and calibrated using the accompanying scripts. The observation setups are described in Table \ref{tab:settings}. The $^{13}$CO data from each program were phase self-calibrated with solutions obtained from the line-free channels within the same spectral window. The continuum was then subtracted from the line emission in the visibility plane using the \texttt{uvcontsub} task. For the sake of comparison with the $^{12}$CO channel maps, the \texttt{mstransform} task was used to average and regrid the spectral line visibilities.  

\begin{figure*}[htp]
\centering
\includegraphics[scale = 0.79]{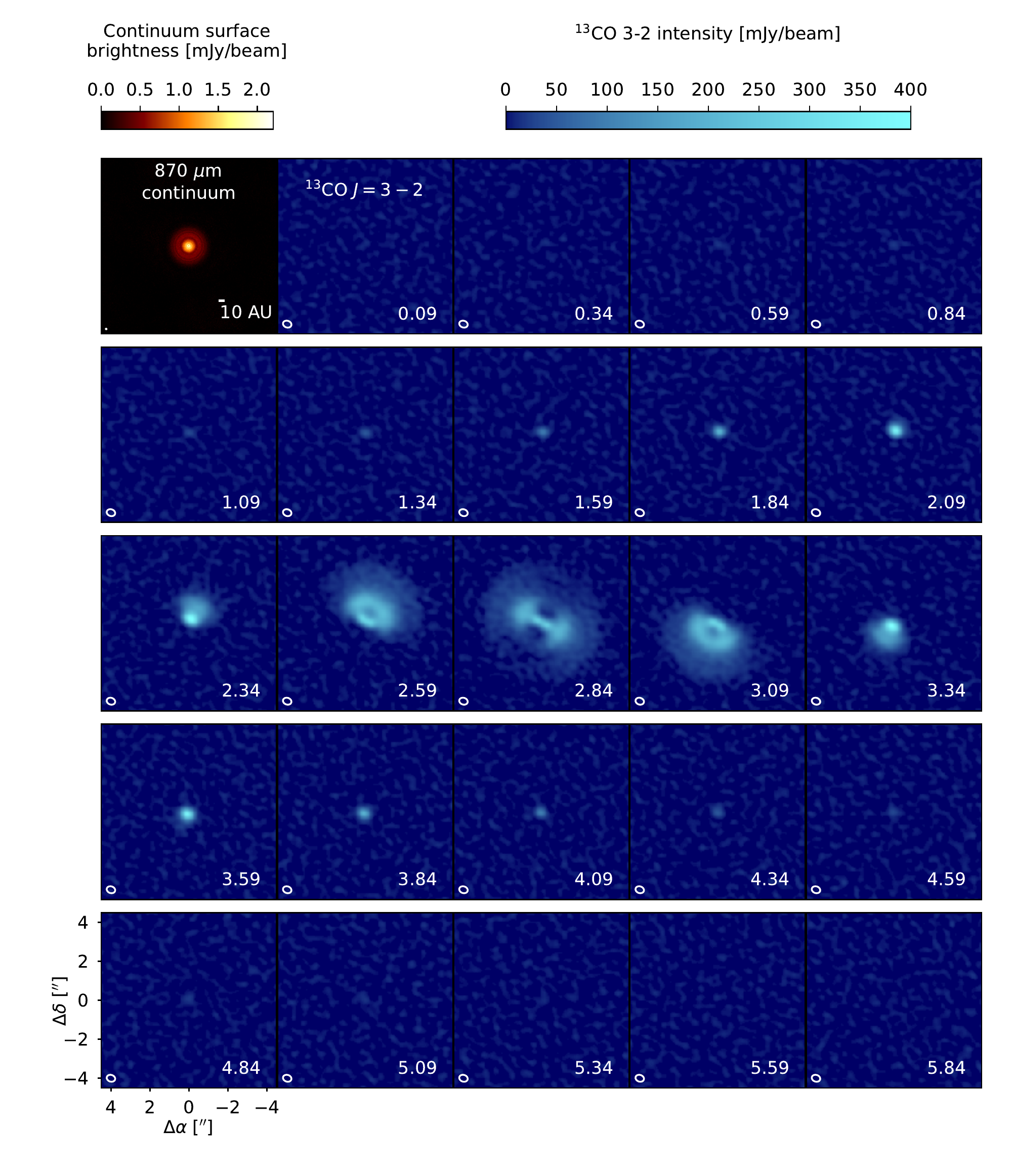}
\caption{Channel maps of the $^{13}$CO $J=3-2$ transition in the TW Hya disk, along with the 870 $\mu$m continuum emission in the upper leftmost panel shown on the same spatial scale. Synthesized beams are drawn in the lower left corner of each panel. The LSR velocity (km s$^{-1}$) for each $^{13}$CO channel is shown in the lower right corner, matching the velocities shown for $^{12}$CO in Figure \ref{fig:chanmap}.}
\label{fig:13COchanmap}
\end{figure*}

The datasets were imaged together with multi-scale CLEAN, using scales of 0, 0.25, 0.5, 0.75, and 1.25 arcsec with Briggs weighting (robust = 0). The resulting image has a synthesized beam with a FWHM of $0\farcs44\times0\farcs35$ (26 $\times$ 21 AU) at a position angle of 71$\fdg$2. The rms measured in nearby signal-free channels is $\approx$ 3.5 mJy beam$^{-1}$. As with the $^{12}$CO data, a primary beam correction was applied to the image cube, shown in Figure \ref{fig:13COchanmap}. 

\end{document}